\documentclass[preprint,amssymb,amsmath,aip,cha,twocolom]{revtex4-1}
\usepackage{graphicx}
\usepackage[caption = false]{subfig}
\usepackage{color}
\usepackage{epsfig}
\usepackage{ifpdf}
\usepackage{url}%
\usepackage{cleveref}
\usepackage{bm}
\usepackage[colorlinks=true,linkcolor=blue]{hyperref}%
\expandafter\ifx\csname package@font\endcsname\relax\else
 \expandafter\expandafter
 \expandafter\usepackage
 \expandafter\expandafter
 \expandafter{\csname package@font\endcsname}%
\fi
\begin{document}
\title{Propagation characteristics of dust$-$acoustic waves in presence of a floating cylindrical object in the DC discharge plasma.}
\author{Mangilal Choudhary}
\email{mangilal@ipr.res.in}
\author{S. Mukherjee}
\author{P. Bandyopadhyay}
\affiliation{Institute for Plasma Research, Bhat, Gandhinagar, 382428, India}
%
\begin{abstract}
The experimental observation of the self$-$excited dust acoustic waves (DAWs) and its propagation characteristics in the absence and presence of a floating cylindrical object are investigated. The experiments are carried out in a direct current (DC) glow discharge dusty plasma in the background of argon gas. Dust particles are found levitated at the interface of plasma and cathode sheath region. The DAWs are spontaneously excited in the dust medium and found to propagate in the direction of ion drift (along the gravity) above a threshold discharge current at lower pressure. The excitation of such low frequency wave is a result of  the ion--dust streaming instability in the dust cloud. The characteristics of the propagating dust acoustic wave get modified in presence of a floating cylindrical object of radius larger than the dust Debye length. Instead of propagating in the vertical direction, the DAWs are found to propagate obliquely in presence of the floating object (kept either vertically or horizontally). In addition, horizontally aligned floating object gives rise to form a wave structure in the cone shaped dust cloud in the sheath region. The change in the propagation characteristics of DAWs are explained on the basis of modified potential (or electric field) distribution, which is a consequence of coupling of sheaths formed around the cylindrical object and the cathode.
\end{abstract}
\maketitle
\section{Introduction}
Dusty plasma is a complex system which is consisted of micron or sub--micron 
sized dust grains in the background of conventional two-component plasma. These 
particles are either formed as a result of agglomeration of reactive species or 
deliberately injected into the plasma. In low temperature plasma environment, 
highly mobile electrons resides on the dust surface and make these dust grains 
negatively charged up to $10^3-10^5$ times of an electronic 
charge\citep{Charging}. Above a critical density, these highly charged dust 
grains inside a plasma show collective nature similar to other plasma species 
(electrons and ions). Theoretical predication as well as experimental 
observation shows that the dusty plasma medium frequently supports the 
excitation of linear and nonlinear waves. Dust acoustic waves  
\citep{raodaw1,daw2,daw3}, dust density waves \citep{ddw1,ddw2,ddw3}, dust 
acoustic transverse waves  \citep{lmode,tsw}, dust lattice waves 
\citep{dlw1,dlw2,tdlw} are some of the linear modes whereas dust acoustic 
solitary waves \cite{kdv,pdasw}, dust ion acoustic shock waves \citep{diasw} 
and dust acoustic shock waves \citep{dasw,expdasw,exp1dasw} are the nonlinear 
modes which have been studied broadly since last couple of decades.\par
After the invention of dust acoustic wave in 1995 \citep{daw2}, an extensive work (see the review paper by Merlino et al. \cite{dawmerlino}) on spontaneously excited DAWs has been carried out both theoretically and experimentally. Dust acoustic mode is an extremely low frequency compressive mode in which the dust particles provide the inertia and the shielding electron and ion clouds provide the thermal pressure effects. To understand the physics underlying the propagation characteristics of such low frequency modes, various experimental configurations have been employed to confine the dust particles \citep{dustydevicer, mddw, ddwfortov, pramanikddw, defectddw, icpddw}. It is experimentally shown in most of the studies that the self$-$excited dust acoustic waves in DC discharge dusty plasma are the consequence either of the streaming ions through dust cloud \citep{instability1,instability2} or dust charge variation in the confined dust cloud \citep{icpddw}. Theoretical studies \citep{instability3,instability4,instability5} confirms that the spontaneous DAW gets triggered in most of the cases either by ions streaming or dust-charge fluctuation in the dust cloud which essentially supports the experimental observation.
\par
The influence of a floating or biased object (more specifically the probe) on 
the confined dust cloud is an research area of current interest. In general, it 
is observed that a dust free region (called void) is formed around a floating 
object inserted inside a dust cloud in an anodic plasma \citep{moveprobe}. The 
dust void around a negatively biased probe and its size dependence on the bias 
voltage is examined by Thomas et al.\citep{void}. In the co-generated dusty 
plasma, a dust free region is formed around a ring shaped positively biased 
electrode\citep{ringvoid}. Klindworth et al.\citep{microgravityvoid} studied the 
dust void around a Langmuir probe in a RF produced dusty plasma under 
microgravity conditions. Kim et al.\citep{diffraction} experimentally 
investigated the dust acoustic wave which gets diffracted during the interaction 
with a floating cylinder. The experimental observation of the interaction of 
flowing dusty plasma with an electrically biased wire is studied by Meyer et 
al.\citep{flowing}. In their study they found the floating wire excites the bow 
like shock structures. The effect of positively (or negatively) biased 
cylindrical probe on a strongly coupled dusty plasma (in crystalline state) is 
reported by Law et al.\citep{circulation}. They observed that the biased probe, 
placed near the RF-sheath edge, alters the crystal properties around it and as a 
result a circulation of the dust particles is observed. { The modification in 
the propagation characteristics of DAWs in presence of a floating object has not 
been explored much in these previous studies and is the subject matter of the 
present manuscript.}
\par
In this paper, we report an experimental observation of self$-$excited dust acoustic wave (DAWs) and its propagation characteristics in absence and presence of a floating cylindrical object in a DC glow discharge plasma. Similar to earlier observations, in absence of the rod, the dust cloud shows a spontaneous excitation of dust acoustic waves for a given discharge condition. The propagation characteristics of the dust acoustic waves are found to get significantly modified during the interaction with the floating rod. The modified characteristics of DAWs are explained on the basis of combined sheath electric field formed due to the presence of a floating object in the cathode sheath. 
\par
The manuscript is organized as follows: Sec.~\ref{sec:exp_setup} deals with the detailed description on the experimental setup. Sec.~\ref{sec:plasma} describes the plasma production and its characterization.  A detailed description of dusty plasma production and its characterization is provided in Sec.~\ref{sec:dusty_plasma}. The experimental observation of dust acoustic waves and its characteristics is discussed in Sec.~\ref{sec:DAW_char}. Sec.~\ref{sec:DAW_rod} highlights the detailed characteristics of DAWs in presence of a floating rod. A brief summary of the work along with a  concluding remarks is provided in Sec.~\ref{sec:conclusion}.
\section{Experimental Setup}\label{sec:exp_setup}
\begin{figure}[h]
\includegraphics[scale=0.60]{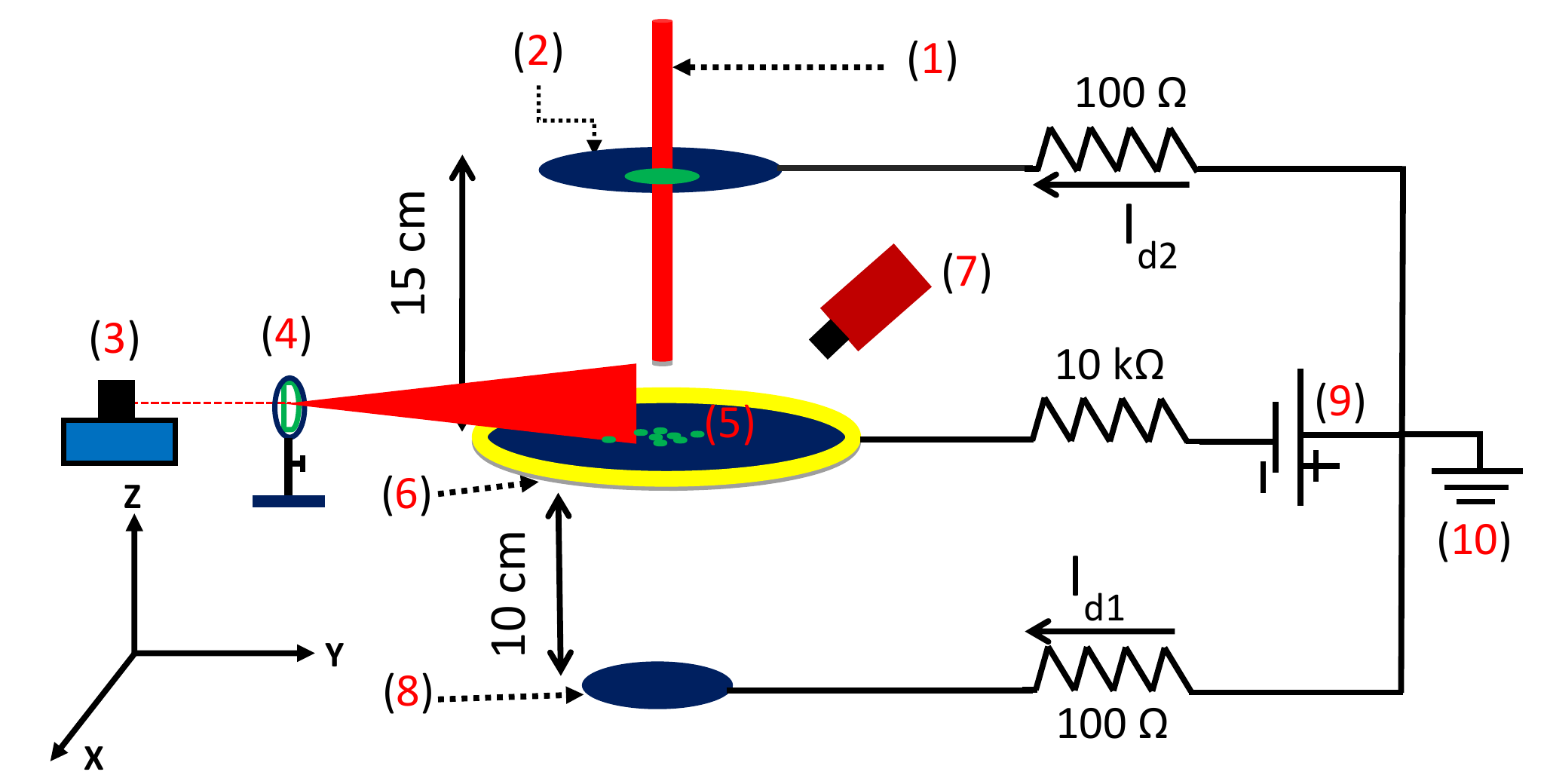}
  \caption{\label{fig:fig1}Schematic of experimental configuration:(1) cylindrical rod. (2) Stainless steel(SS) flange (working as an anode). (3) solid$-$state red laser. (4) plano$-$convex cylindrical lens. (5) kaolin dust particles.(6) SS disk electrode (cathode) with a ring like step at corner (represented by yellow ring). (7) CCD camera. (8) Stainless Steel circular electrode (anode). (9) Discharge power supply (1kV and 3A) . (10) grounded. $I_{d1}$ and $I_{d2}$ represent the currents flowing in separate discharge path.}
 \end{figure}

Experiments are performed in a borosilicate glass tube with an inner diameter of 
15 $cm$ and length of 60 $cm$. The glass tube has 5 radial and 2 axial ports, 
used for pumping, gas feeding, plasma production and plasma/dusty plasma 
diagnostics purposes. A rotary pump and gas dosing valve are attached to a 
Stainless Steel (SS) buffer chamber of 30 $cm$ length and 15 $cm$ diameter, 
which is connected to the glass tube.{ The geometrical (3D) view of the 
experimental setup is described in more details elsewhere \citep{expsystem}}. 
This experimental assembly prevents the direct gas flow induce neutral drag on 
the dust particles in the main experimental chamber where the dusty plasma 
experiments are carried out. The schematic of operating configuration is 
presented in the Fig.~\ref{fig:fig1}. A SS cylindrical rod of 5 $mm$ diameter is 
inserted in the experimental chamber over the cathode using a Wilson 
feedthrough. The experimental chamber is evacuated below $10^{-3}\ mbar$ 
pressure using the rotary pump and the argon gas is then fed into the chamber. 
The chamber is pumped down again to the base pressure. This process is repeated 
several times to reduce the impurities from the vacuum vessel. Finally the 
operating pressure is set to 0.07 $mbar$ by adjusting the gas dosing valve and 
pumping speed.
 \section{Plasma production and its characterization} \label{sec:plasma} 
 A DC glow discharge plasma is initiated by applying a voltage in the range of 
300--500 V in between a  cathode and two anodes as shown in Fig.~\ref{fig:fig1}. 
The disk shaped cathode of 10 $cm$ diameter is having plane surface in one side 
whereas the other side having a step of 5 $mm$ width and 2 $mm$ height around 
its periphery. The plane surface of the cathode is parallel to a SS disk of 4 
$cm$ in diameter, which is a one of the anodes (lower anode). This lower anode 
is kept 10 $cm$ below from the cathode. Another anode (upper anode) is a SS 
flange of 7 $cm$ diameter is parallel to the dust particles containing cathode 
surface. This upper anode is fixed 15 $cm$ above the cathode. {These above 
configuration of anodes and cathode demands minimum voltage to breakdown the 
argon gas}. Both the anodes are kept grounded through a resistance of $R=100 
\Omega$ to measure the discharge currents in each path. The experimental chamber 
is purposefully made of dielectric material and all the axial and radial ports 
are closed by toughened glasses and/or perspex flanges so that additional 
discharge paths can be avoided. There are some advantages to use this discharge 
configuration for performing dusty plasma experiments. In this present set of 
experiments, the dust particles are sprinkled homogeneously over the cathode 
surface which acts as a dielectric covered electrode. For breaking down the gas, 
cathode electrode is made negatively biased with respect to the grounded anodes. 
In absence of lower anode, 30$-$40 V more bias voltage is required to break down 
the gas and as a result it becomes difficult to form an appropriate dust cloud 
at lower gas pressure.
 \par
\begin{figure}[h]
 \centering
\subfloat{{\includegraphics[scale=0.4]{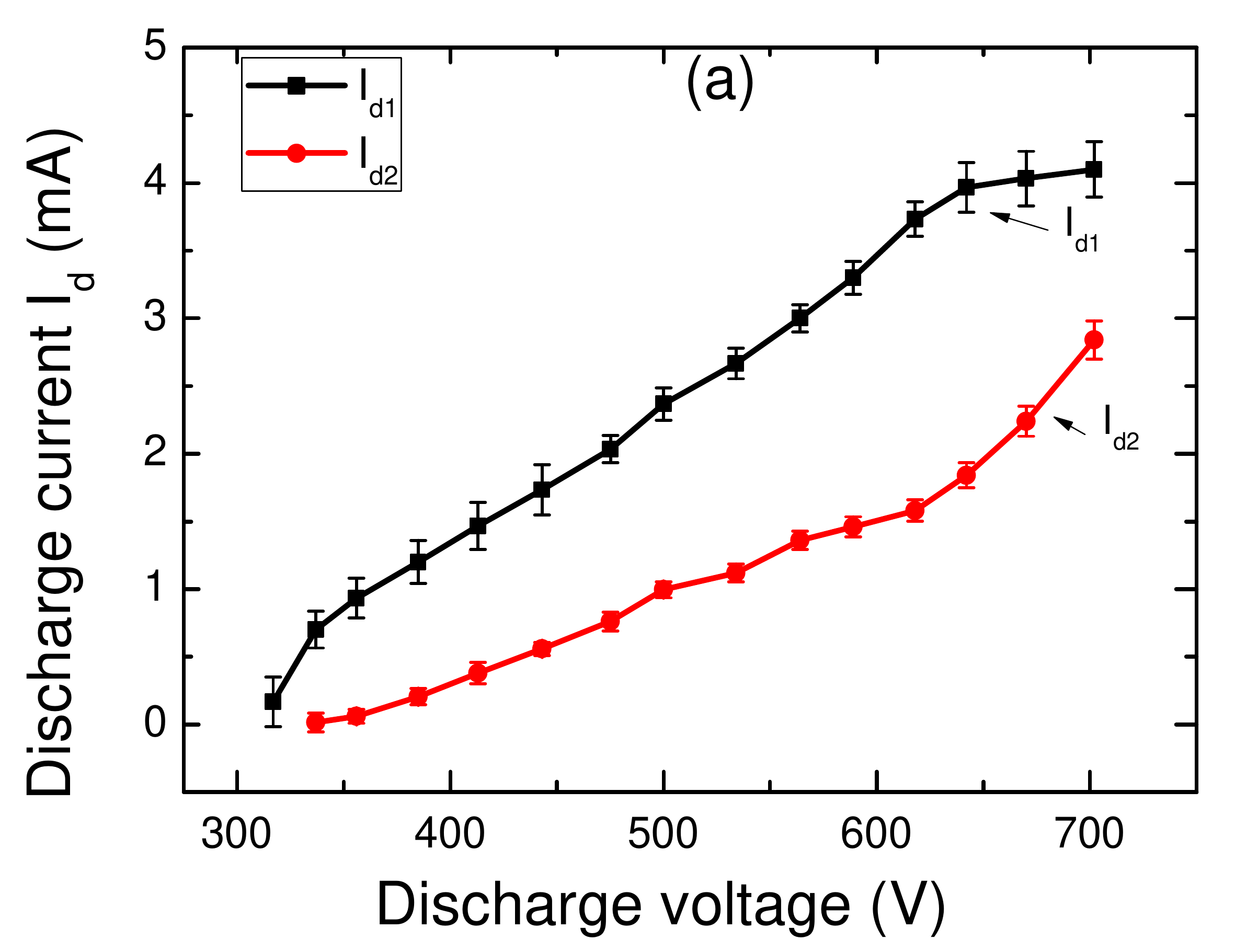}}}%
\vspace*{-0.13in}
  \qquad
    \subfloat{{\includegraphics[scale=0.4]{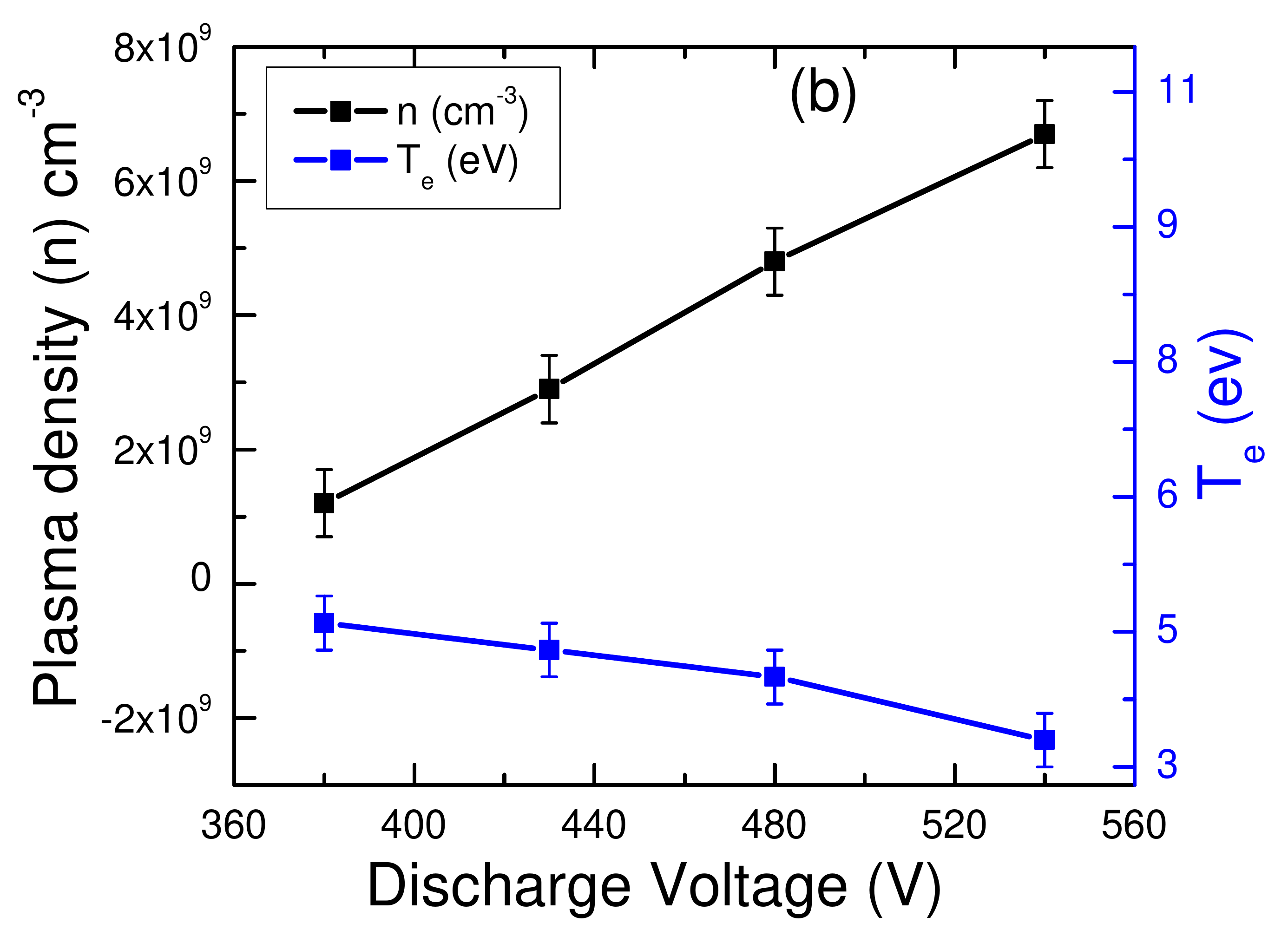}}}%
  \caption{\label{fig:fig2}(a) Discharge current $(I_{d1}$ and $I_{d2})$ variation with discharge voltage (between cathode and anodes). Discharge currents are computed by dividing the voltage drop across a resistance of 100 $\Omega$. (b) Variation of plasma density $(n)$ and electron temperature $(T_e)$ with the discharge voltage. The argon pressure is set to 0.07 $mbar$}
  \end{figure}

The variation of discharge current $(I_d)$ in both the discharge paths with  discharge voltage $(V_d)$ are measured in absence of dust particle. The  $I_d-V_d$ curves at pressure p = 0.07 $mbar$ are displayed in 
Fig.~\ref{fig:fig2}(a). It is clearly seen in Fig.~\ref{fig:fig2}(a) that at  first, $I_{d1}$ (represented by closed rectangular) starts flowing at 320 V whereas $I_{d2}$ (represented by closed circle) is initiated at 340 V. It  indicates that the plasma initially forms in between the cathode and lower anode at about 320 V and later it strikes between the cathode and the upper  anode. It is also to be noted that $I_{d1}$ stays always higher than the 
$I_{d2}$ for the whole range of discharge voltage. {It is also worth mentioning 
that at a particular discharge condition, the flowing currents ($I_{d1}$ and 
$I_{d2}$) in each path depend upon the dimension of each electrodes.}  
\par
 The plasma parameters at various discharge conditions are measured using a 
cylindrical Langmuir probe of length 7 $mm$ and radius of 0.25 $mm$. { The 
variations of plasma parameters with discharge voltage are displayed in the 
Fig.~\ref{fig:fig2}(b). The plasma density ($n$) and the electron temperature 
($T_e$) change from $\sim$ 1$\times10^9$ to 7$\times10^9$ $cm^{-3}$ and $\sim$ 
2.5 to 5 eV, respectively with the change of discharge voltage from 380 to 540 V 
at a constant pressure p = 0.07 $mbar$. The measured values shows that the 
plasma density increases whereas the electron temperature decreases with the 
increase of discharge voltage. This essentially signifies that with the increase 
of discharge voltage the electron Debye length 
($\lambda_e=\sqrt{\epsilon_0kT_e/ne^2}$) decreases}. The ions are assumed at 
room temperature i.e.$T_i\sim 0.025$ eV. At a given discharge condition, these 
plasma parameters (measured outside the cathode sheath region) are used to 
calculate the dusty plasma parameters.
 \section{Dusty plasma and its characterization}\label{sec:dusty_plasma}
In course of dusty plasma experiments, kaolin dust particles with mass density $(\rho)$ of 2.6 $gm/{cm^3}$ of radii $(r_d)$ ranging from 0.5 to 5 $\mu$m, are sprinkled over the cathode surface. For producing the dusty plasma, DC glow discharge is initiated between cathode and anodes at 0.07 $mbar$. The discharge current$-$voltage characteristics in presence of dust particles follow the similar behavior as shown in Fig~\ref{fig:fig2} but with slightly lower values of discharge currents. In presence of the plasma, the dust particles get negatively charged because of impinging of more electrons than ions on the surface of dust particles. The negatively charged dust particles lifted up from the cathode surface and levitate at the interface of plasma and cathode sheath region. The cathode geometry (ring at its periphery)provides the better radial confinement to the levitated dust particles. The levitated dust cloud is then illuminated by a red laser of 632 $nm$ wavelengths and power ranges from 1$-$100 $mW$. The point laser source is converted into a laser sheet using a plano--convex cylindrical lens of focal length of 25 $mm$ which can be oriented in vertical as well as horizontal plane. This laser sheet is efficient enough to illuminate $\sim$2 $mm$ wide vertical (or horizontal) slice of the dust cloud. Due to the poly--dispersive nature of the particles, the heavier dust particles levitate at the bottom whereas the lighter particle levitate at the top of the sheath by balancing the electrostatic force and the gravitational force. The dynamics of the particles in vertical as well as in horizontal plane are recorded at 130 frames per second (fps) by a Lumenera CCD camera with the resolution 1088 pixels$\times$2048 pixels. The data is then transfer to a high speed computer for the purpose of further analysis using different softwares.\par
\begin{figure}[h]
 \centering
   \includegraphics[scale=0.40]{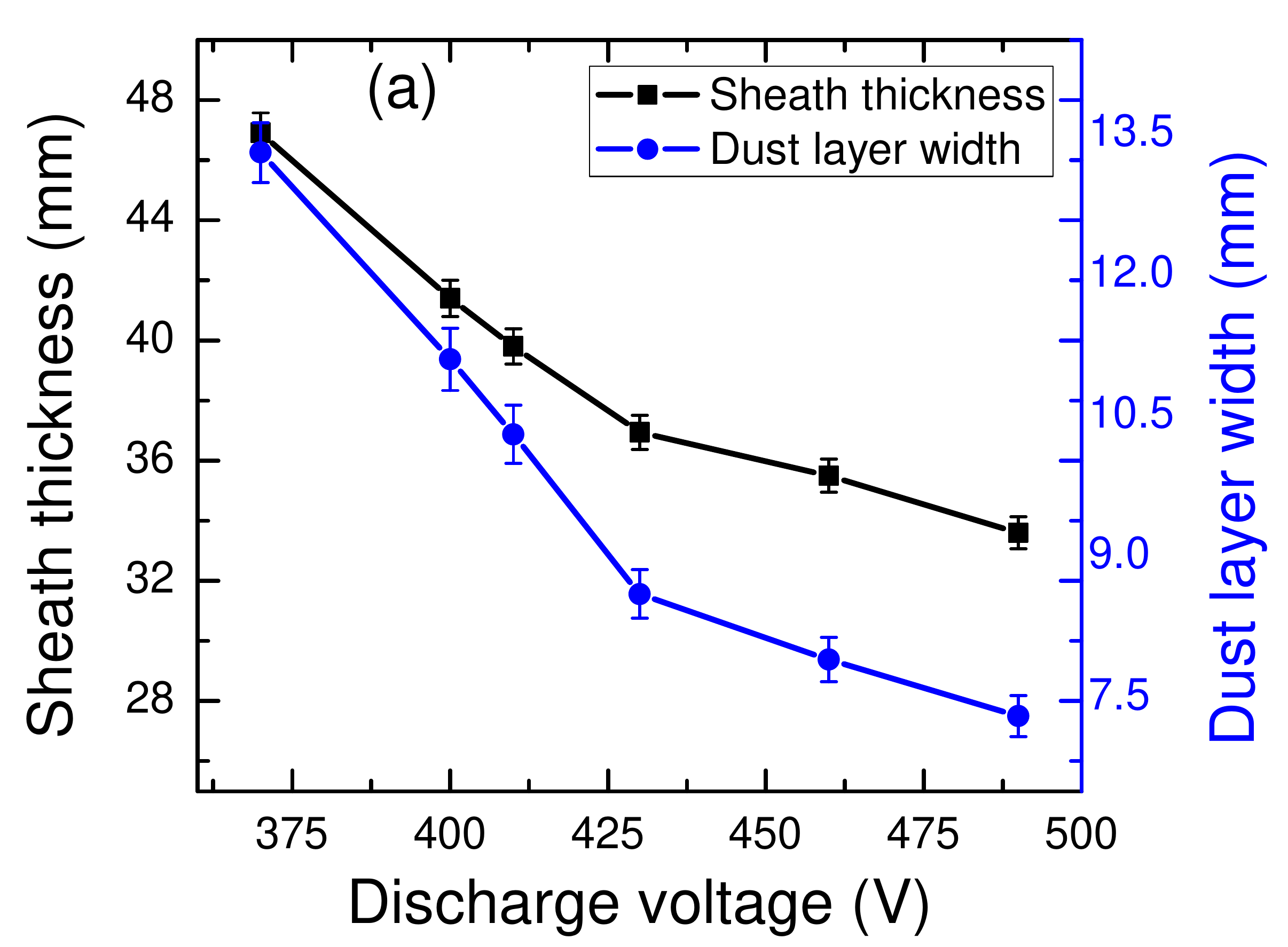}
\caption{\label{fig:fig3}The variation of sheath thickness (solid rectangular) and dust cloud width (solid sphere) with applied discharge voltages. The argon filling pressure during the experiment is set at 0.07 $mbar$.}
  \end{figure}

 The effect of discharge parameters on the levitated dust cloud is also 
necessary to understand the propagation characteristics of dust acoustic waves. 
Therefore the sheath thickness and the width of the dust cloud are measured at 
given pressure for various discharge voltages. The sheath thickness is estimated 
as the distance between the cathode and the topmost layer of the dust particles 
whereas the difference of topmost and the bottommost layers gives the width of 
the dust cloud. The sheath thickness is found to decrease with the increase of 
the applied discharge voltage as shown in Fig.~\ref{fig:fig3} (represented by 
solid rectangular). { This experimental observations is in accordance with the 
theoretical estimation. According to Child-Langmuir law, the high voltage 
cathode sheath is function of the plasma Debye length (in general sheath 
thickness is ten to hundred times more than the Debye length 
\citep{sheaththickness}). As mentioned earlier, with the increase of discharge 
voltage the electron Debye length decreases which results in the decrease of 
sheath thickness. As sheath thickness reduces, the electric field gradient at 
sheath edge increases therefore the dust cloud  get suppressed.} The variation 
of width of the dust cloud with discharge voltage is also shown in 
Fig.~\ref{fig:fig3} (represented by solid sphere).
 \section{Excitation and propagation characteristics of Dust Acoustic Waves }\label{sec:DAW_char}
\begin{figure}[h]
 \centering
\subfloat{{\includegraphics[scale=0.61]{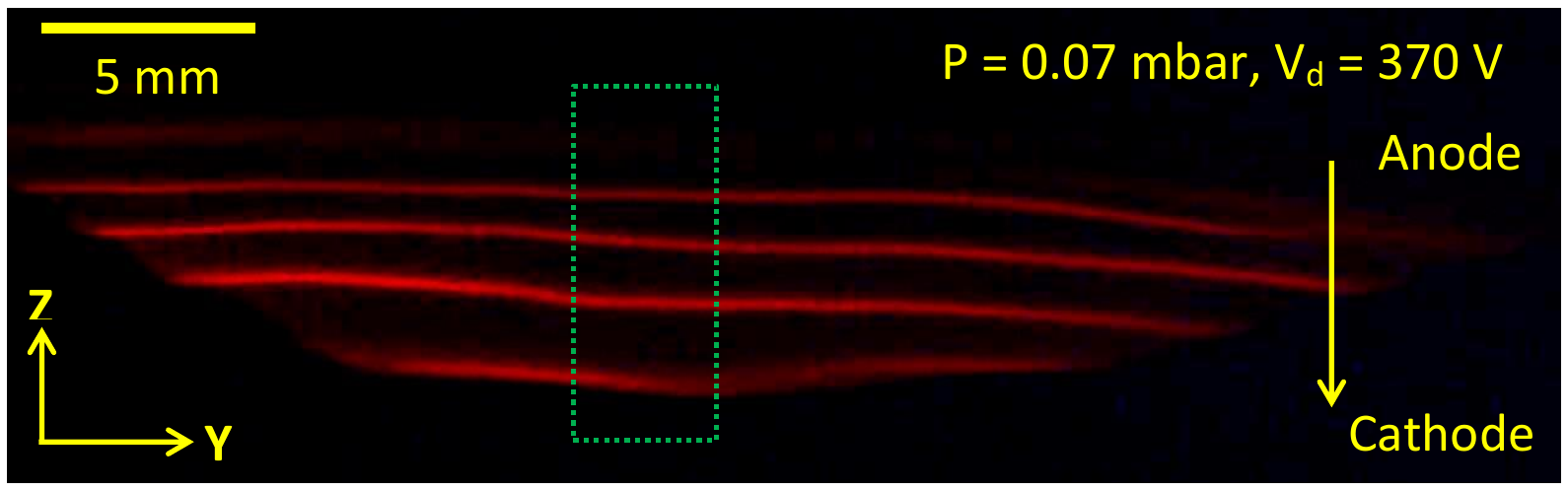}}}%
\vspace*{-0.13in}
  \qquad
    \subfloat{{\includegraphics[scale=0.72]{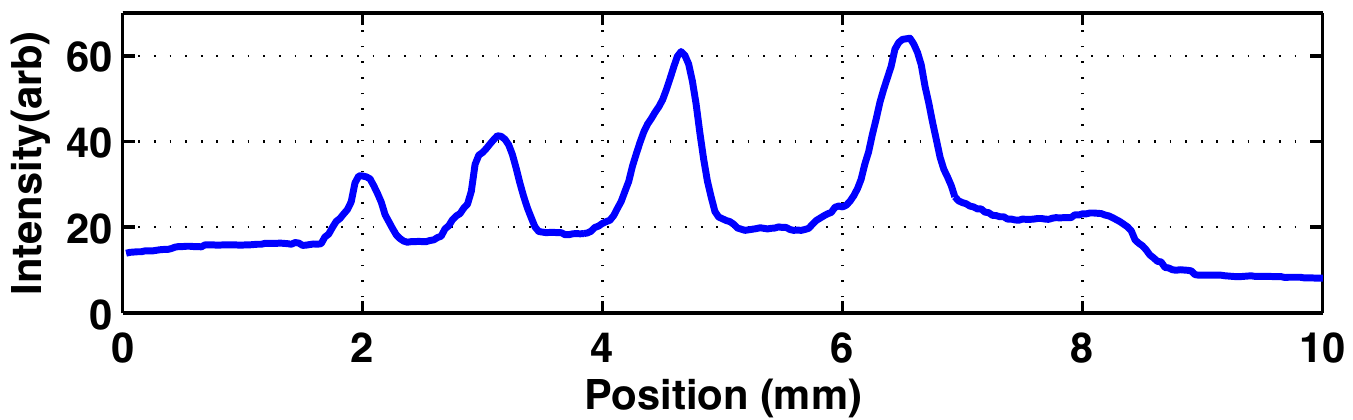}}}%
  \caption{\label{fig:fig4}(a) A video image of Dust Acoustic Waves in vertical (Y$-$Z) plane. Arrow indicates the velocity vector for DAW from anode to cathode. (b) Average intensity profile of scattered laser light coming from dust particles (in dotted region) in -$\hat{z}$ direction. Discharge voltage and gas pressure are set at 370 V and 0.07 $mbar$, respectively.}
  \end{figure}

\begin{figure}[h]
 \centering
{\includegraphics[scale=0.80]{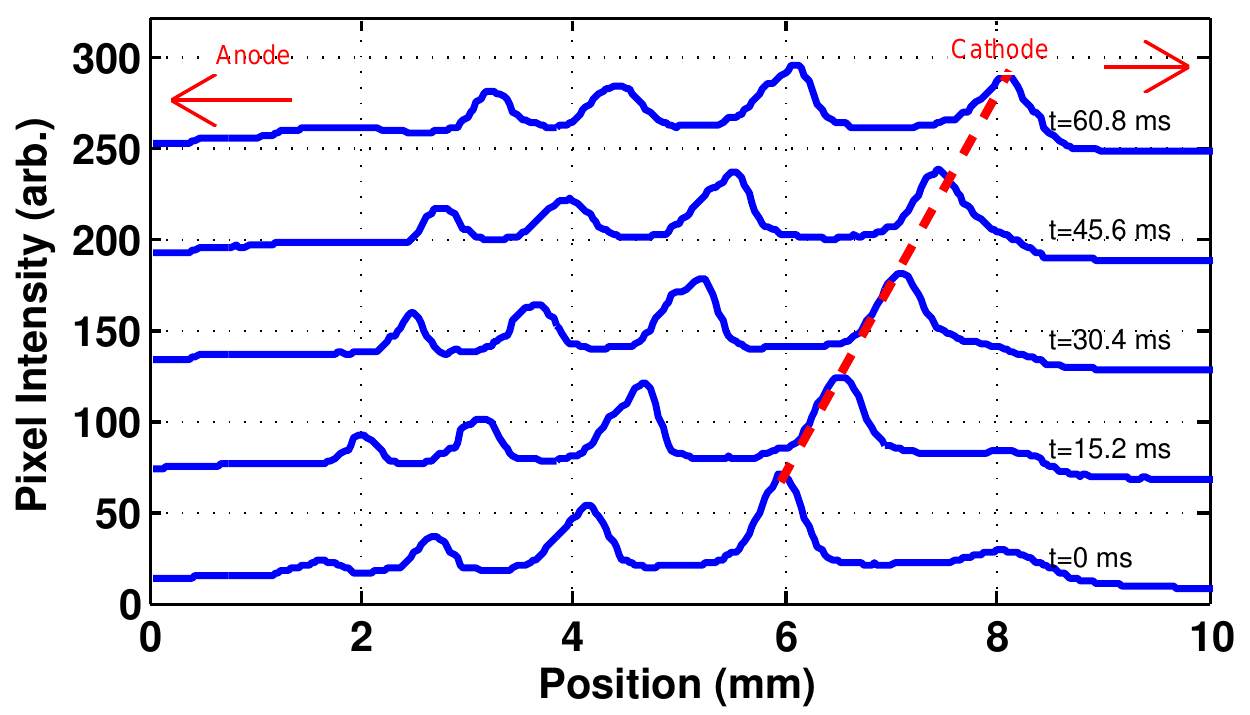}}%
\caption{\label{fig:fig5}Time evolution of intensity profile of five video frames taken at time step of 15.2 $ms$ along the selected region (in Fig.~\ref{fig:fig4}(a)) in the vertical plane. Discharge voltage and argon pressure are set at 370 V and 0.07 $mbar$ respectively for this particular set of experiments}
  \end{figure}

 The dust particles form a steady state equilibrium approximately at 40$-$45 
$mm$ above the cathode at $V_d=340$ V and $p=0.07$ mbar. If the discharge 
voltage is increased slightly to $\sim$ 350 V, an appearance of spontaneous Dust 
Acoustic Wave (DAW) is found. {  This wave originates in the upper part of the 
levitated dust cloud and propagates towards the cathode (along the gravity)}. A 
typical video image of this self$-$excited DAWs (in Y$-$Z plane) at $V_d = 380$ 
V and $p = 0.07$ mbar is presented in Fig.~\ref{fig:fig4}$(a)$. { The 
stable dust cloud shows the excitation of spontaneous dust acoustic wave when 
the ratio $E/p$ crosses a threshold value \citep{instability1}. This ratio 
increases either with the increase of electric filed (or the discharge voltage) 
or decrease of gas pressure. At higher pressure and higher discharge voltage, 
the wave could not excites due to frequent ion-neutral collision. Whereas at 
higher discharge voltage and lower pressure, the dust acoustic wave gets highly 
unstable \cite{pramanikddw}. Hence keeping both the points into mind, a 
constant pressure at p = 0.07 $mbar$ is chosen to perform the experiments at a 
range of discharge voltage }. The average intensity profile of 
Fig.~\ref{fig:fig4}$(a)$ is shown in the Fig.~\ref{fig:fig4}$(b)$. { 
It is observed in the figure that the wave initially gets excited with smaller 
amplitude on the top and then propagates with increasing amplitude. It happens 
due to the non-uniformity of electric field inside the cathode sheath. The ions 
initially enter into the sheath with Bohm velocity and then accelerate towards 
the cathode and as a result these ions exert more pressure on the dust particles 
when they reach towards the cathode. It results in the increase of  amplitude of 
DAWs while they propagate towards the cathodes.}
For the detailed characterization of DAWs, few consecutive frames are 
considered. The time evolution of average intensity profile is shown in the 
Fig.~\ref{fig:fig5}. The red dashed line in the figure indicates the trajectory 
of a particular crest of the DAW. The phase velocity $(v_{ph})$ and wavelength 
$(\lambda)$ are found to be $\sim$ 2$-$3$~cm/sec$ and $\sim$ $1.5-2.5$ $mm$, 
respectively. The wave frequency $(f_d)$ can be estimated from $v_{ph}$ and 
$\lambda$ and comes out to be $\sim$ 10$-$15 Hz. { It is also to be noted that 
the distance between two consecutive crests increases with time as the higher 
(smaller) amplitude wave moves with faster (slower) velocity\citep{pdasw} }. 
These values of the wave parameters are consistent with the theoretical 
estimation of dust acoustic waves. For cold dust ($T_d$ = 0) and in long 
wavelength limit $(k_D>>1)$, the phase velocity of dust acoustic wave $v_{ph} = 
z_d\left({kT_i}{n_{d0}}/{m_d}n_{i0}\right)^{1/2}$ , where $k$, $T_i$, $m_d$, 
$n_{d0}$, $ n_{i0}$, and $z_d$  are Boltzmann constant, ion temperature, dust 
particle mass, equilibrium dust density, ion density and dust charge number, 
respectively\citep{sounddaw}. { $z_d$ can be estimated by assuming 
the dust grains are spherical capacitors. Hence, the charge resides on the dust 
surface can be expressed as  $Q_d = z_de=4 \pi \varepsilon_0 r_d V_s$. The dust 
surface potential, $V_s \sim -4kT_e/e$, is estimated from OML theory 
\citep{Charging} }. The estimated phase velocity ($v_{ph}$) of the DAW is found 
to be $\sim$ $3.3$ $cm/sec$ for parameters $T_e = 4$~eV, $T_i=0.025$ eV, $m_d$ 
(for an average radius $r_d \sim $ 3 $\mu$m) $\sim$ $3\times10^{-13}$ kg, 
$n_{d0}\sim$ 1$\times$ $10^5$ $cm^{-3}$, $n_{i0}\sim$ $1\times$ $10^9$ 
$cm^{-3}$, and $z_d \sim$ 3 $\times$ $10^4$. Therefore, theoretical predicted 
value of the phase velocity is in close agreement with the experimental phase 
velocity of DAWs.\par
As discussed, the spontaneous DAWs are observed to propagate in the direction 
of anode to cathode (along the gravity) which essentially in the direction of 
ion flow. It suggests that the DAW in the present set of experiment could be 
excited due to ion$-$dust streaming instability. According to earlier 
investigations \citep{instability1,instability2}, in a dc discharge dusty 
plasma, the ion$-$dust streaming is responsible to excite DAW if ion$-$drift 
velocity $(u_{i0})$ becomes in the  order or higher than the ion thermal 
velocity $(v_{Ti})$ i.e. $u_{i0}\geq v_{Ti}$. In our experiments, the ion 
thermal speed $v_{Ti} = \sqrt{(8kT_i/m_i\pi)} \sim 3.9\times10^4$ cm/sec, where, 
$m_i$ $(=6.7\times10^{-26}$ kg) is the mass of argon ions. The ion drift 
velocity $(u_{i0})$ can be estimated from, $u_{i0} = \mu_i$E, where $\mu_i$ and 
E are the ion mobility and the sheath electric field, respectively. The sheath 
electric field is estimated from the equilibrium condition of dust particles 
where the gravitational force is exactly balanced by the electrostatic force. 
Hence at equilibrium, $e z_d E = m_d g$, where $e$ $(= 1.6\times 10^{-19}$ C) is 
the electronic charge and $g$ $(= 9.8$ m/sec$^2$) is the acceleration due to 
gravity. For a given values of $m_d \sim$ $3\times10^{-13}$ kg and $z_d \sim$ 3 
$\times 10^4$, the calculated electric field $E\sim$ $6~V/cm$. The mobility of 
Ar ions $\sim 1.8\times 10^3/p(torr)$ $cm^2 s^{-1} V^{-1}$\citep{tsw} . At 
p=0.07 $mbar$, $\mu_i \sim 3.5\times10^4$ $cm^2 s^{-1} V^{-1}$. Therefore, the 
estimated ion$-$drift velocity, $u_{i0}$, becomes $\sim 2 \times 10^5$ $cm/sec$ 
({for E~$\sim$6 V/cm}). The ratio of $u_{i0}/v_{Ti}$ is $\sim 5$. Hence the 
above estimation assures that the excitation of DAWs in our experiments is 
mainly due to the ion--streaming through the dust cloud. It is also found that 
at higher pressure $(\geq 0.2~mbar)$, the ion$-$neutral collision frequency 
increases as a result the oscillations of dust particles get suppressed due to 
frequent ion$-$neutral collisions and therefore the waves could not be excited.
\section{Characteristics of Dust Acoustic Waves in presence of a floating rod}\label{sec:DAW_rod}
 \subsection{Vertically aligned floating rod}
After a thorough characterization of Dust Acoustic Wave (DAW), an experiment is 
carried out to study the modification of its propagation characteristics in 
presence of a floating object. In this set of experiments, a cylindrical rod is 
chosen as a floating object and introduced inside the plasma perpendicular to 
the cathode plane as mentioned in Sec.~\ref{sec:exp_setup}. The dimension of the 
rod is taken in such a way that $r  >> \lambda_D$. where, $r$ is the 
radius of the rod and $\lambda_D$ $\sim$ $\lambda_{Di}$ ($= \sqrt{\epsilon_0 
kT_i/ne^2}$) is the dust Debye length. In our discharge regimes, $\lambda_D$ 
varies from $0.015$ $mm$ to $0.03$ $mm$. In this regime, it is 
found that a wire of radius, $r\sim \lambda_D$, does not affect (perturb) the 
dust cloud therefore radius of the rod is chosen bigger in 
dimension (($r$ = 2.5~$mm$)) so 
that its influence on DAW can be observed. The effective changes on the dust 
cloud or DAW in presence of floating cylindrical rod near the upper layer of 
dust cloud is examined for three different discharge voltages at a particular 
pressure, p = 0.07 $mbar$. The dust free region (void) around the floating rod 
is found to be observed at lower discharge voltage where the dust cloud does not 
exhibit wave structures. This is the case when discharge voltage is $\sim 350$ V 
at the pressure 0.07 $mbar$. However, the detail study on dust free region 
(void) around the floating rod is not the scope of present study and will be 
reported in future.\par
  \begin{figure}[h]
 \centering
\includegraphics[scale=0.74]{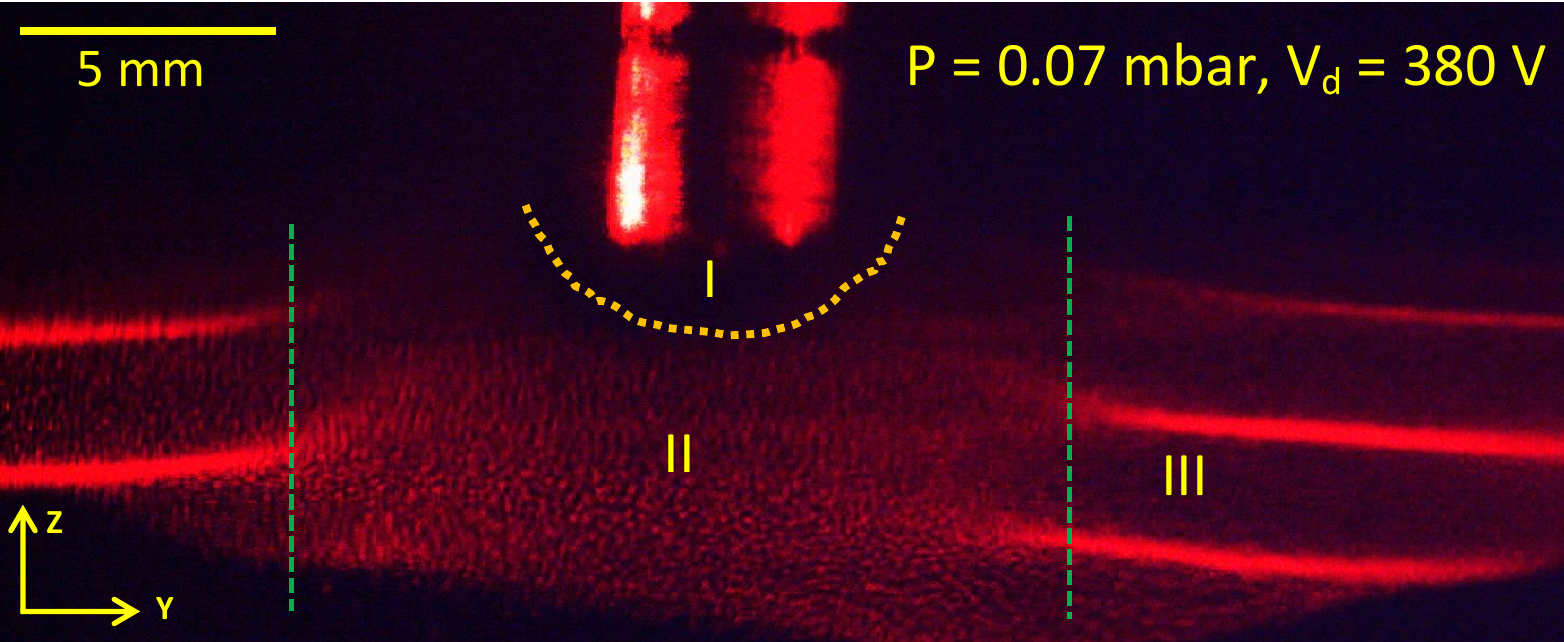} 
 \caption{\label{fig:fig6}A video image of dust acoustic waves in presence of a vertically aligned floating rod. Three distinct regions are observed: (I) Dust void (dotted curve). (II) Stationary dust cloud (region between dotted curve and lines). (III). dust acoustic wave (outside the dotted lines). Argon pressure and discharge voltage are set at 0.07 $mbar$ and 380 V, respectively.}
  \end{figure}
At higher discharge voltage (beyond $380$ V) and the presence of the floating rod (kept far from the dust cloud), the levitated dust cloud exhibits similar kind of propagation of DAWs in the direction of ion flow/gravity as discussed in Sec.~\ref{sec:DAW_char}. 
In this situation, the measured floating potential of the rod which is placed near the upper layer of levitated dust cloud  is $\sim-12$ V. To study the modification of wave properties, the horizontal (in X--Y plane) as well as the vertical (in Y--Z plane) slices of images near the floating rod are captured.  A typical image (see Fig.~\ref{fig:fig6}) of Y--Z plane shows the influence of cylindrical rod on the propagation of DAWs. The consecutive frames are used to analyze the dynamical behavior of the dust grains near by the rod. Three distinct regions are found when the rod is brought near the topmost layer of the dust cloud. 
These regions are classified as: region$-$I, dust free region near the rod 
(between rod and dotted curve), region$-$II, stable dust cloud (between the 
dotted curve and dotted lines) and region$-$III, dust acoustic wave (outside 
the dotted lines). The length of dust free region (length from rod surface to 
dotted curve) is $\sim$ 2 $mm$ whereas the length of affected region (length 
from rod surface to dotted line) is extended upto $\sim 6-7$ $mm$. Outside the 
dust free region (in region$-$II), the dust grains only show the random motion 
instead of participating in the propagation of DAW. The boundary of different 
regions changes with the change of discharge parameters and the dimension of 
the rod.\par 
In the presence of a floating rod, the sheath around it modifies the  
electric field profile of cathode sheath. The resultant electric field is the 
consequence of the coupling between the sheaths formed around the rod and the 
cathode, which has been investigated in details by Barnat \textit{et al.} 
\cite{Efieldaroundprobe} in a two component plasma. Hence, as a result the 
dynamics of ions as well as dust particles get modified in the overlapping 
sheath region. In the region--I, dust grains are expelled near the rod surface 
due to the strong sheath electric field which cause the formation of dust void 
\citep{moveprobe,void}. The random motion of dust grains in the region--II is 
result of the suppression of ion streaming towards the cathode (along the 
direction of gravity). It is observed that the instabilities in dust cloud are 
triggered above a critical electric field \citep{instability3}.  {  
In presence of the rod, the electric field becomes weaker 
\cite{Efieldaroundprobe} (in the region --II) that causes the suppression of 
ion--streaming instabilities in the dust cloud and transforms the wave crests 
into stable dust cloud}. In the region--III, the influence of floating rod is 
negligible therefore the dust cloud exhibits the usual dust acoustic waves.   
\par
\begin{figure}[h]
 \centering
\subfloat{{\includegraphics[scale=0.65]{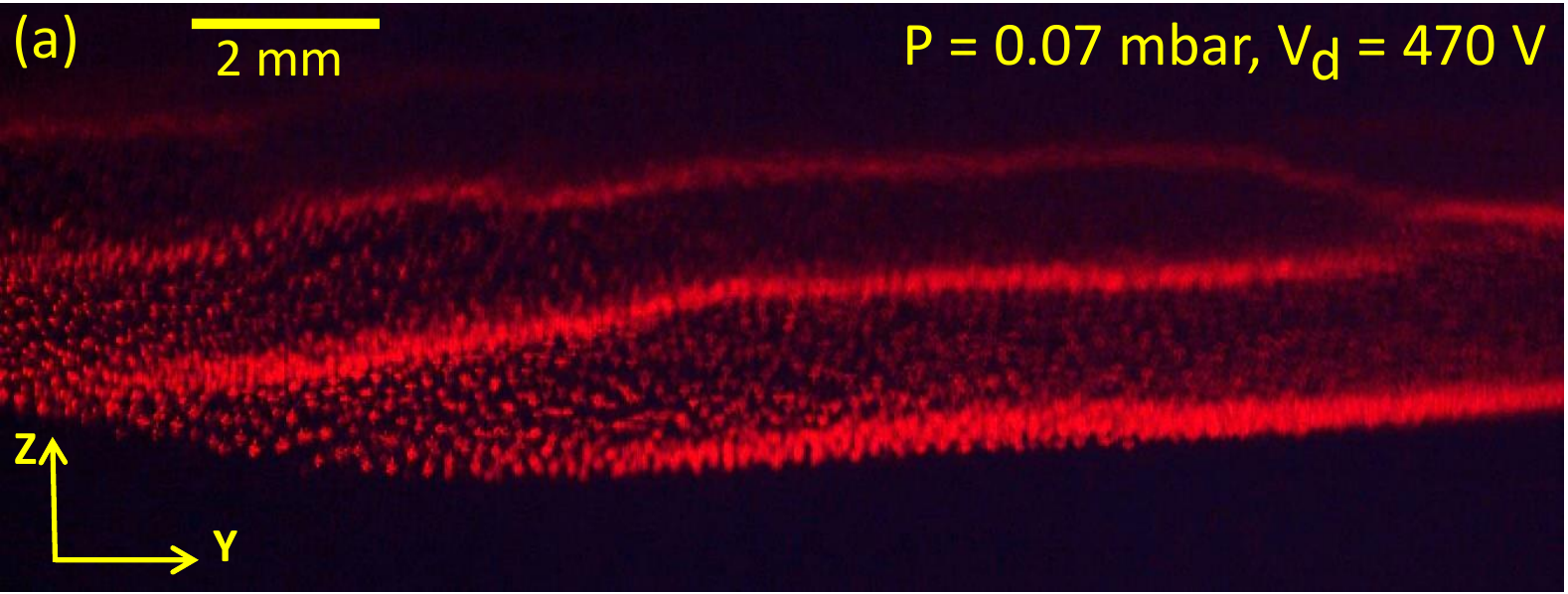} }}%
\vspace*{-0.13in}
   \qquad
   \subfloat{{\includegraphics[scale=0.65]{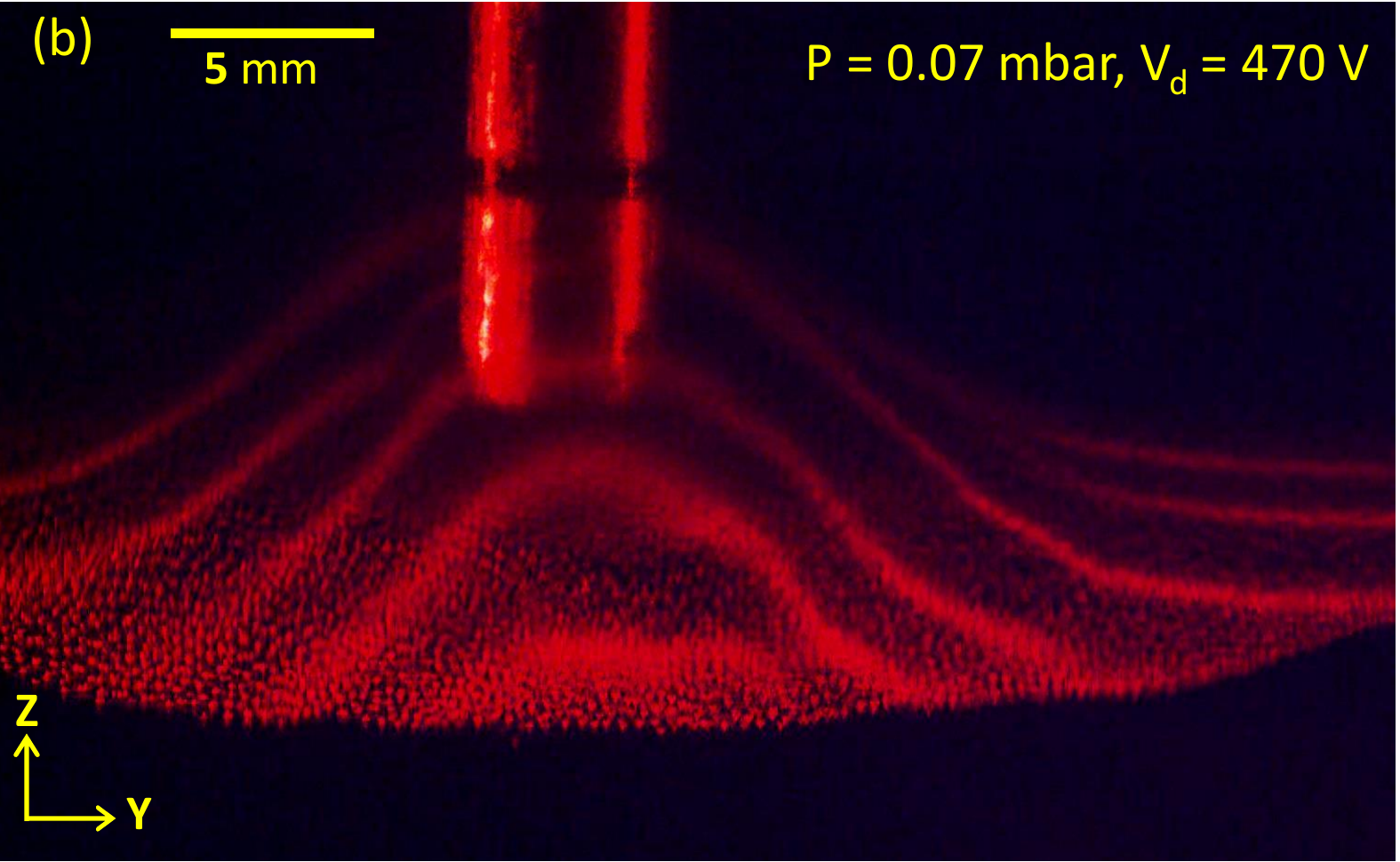} }}%
    \vspace*{-0.14in}
    \qquad
   \subfloat{{\includegraphics[scale=0.748]{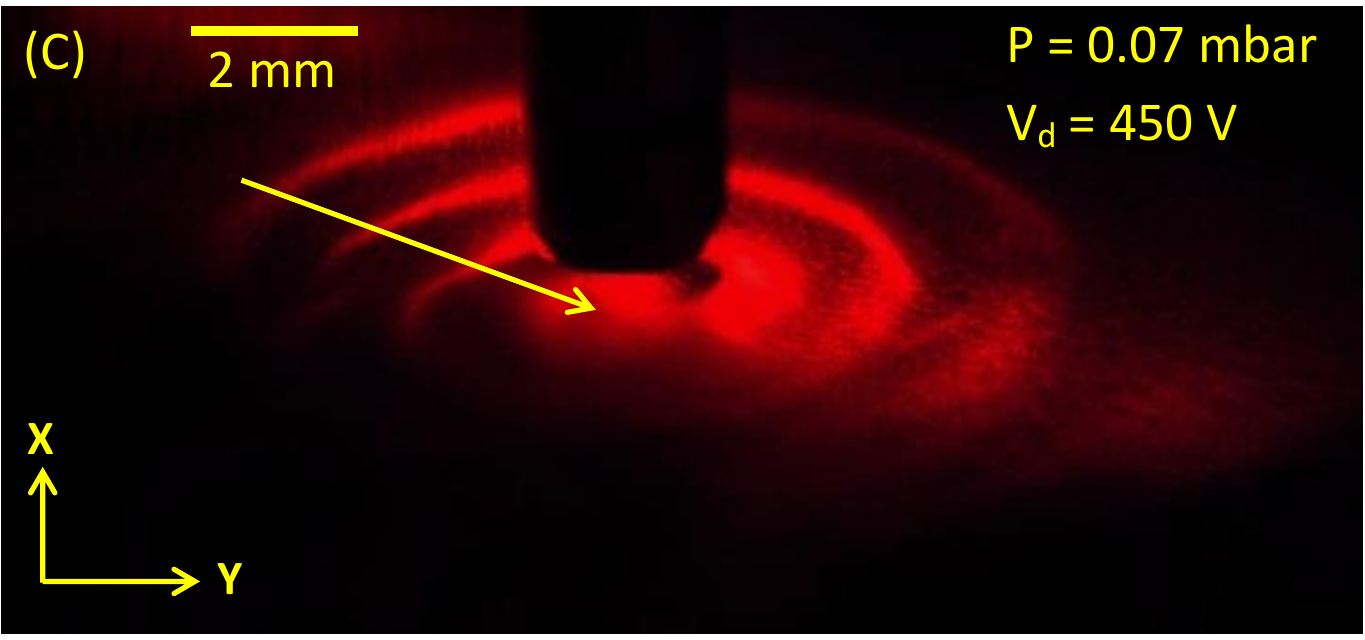} }}%
   \caption{\label{fig:fig7}(a) A video image of dust$-$acoustic waves in vertical plane (in absence of rod).(b) Image of modified DAWs in vertical (Y$-$Z) plane with floating rod where discharge voltage is 470 V. (c) A video image of DAWs in horizontal (radial plane of rod) plane just below the floating rod. Arrow indicates the velocity vector of DAWs in this plane. Argon filling pressure and discharge voltage are set at 0.07 $mbar$ and 450 V, respectively}.
  \end{figure}

With the further increase of discharge voltage (beyond 450 V), the DAWs become unstable (see Fig.~\ref{fig:fig7}(a)) due to the higher value of $E/p$ as reported in ref \citep{instability1} but still propagate in the direction of cathode (along the gravity($-\hat{z}$)). With the increase of discharge voltage, the sheath thickness reduces to nearly 28 $mm$ (as shown in Fig.~\ref{fig:fig3}) whereas the sheath electric field increases. Hence the ratio of $E/p$ increases for a given pressure. In this situation, the average wavelength and velocity of DAW are measured as $\sim$ 2 $mm$ and $\sim$ 4 $cm/sec$, respectively. \par
In this particular discharge condition,  when the floating rod, with potential 
$\sim -8$ V, is moved from plasma to the upper layer of dust cloud, an unique 
features in the dust cloud adjacent to the object is observed. A typical image 
of modified DAW in Y--Z plane $\sim 2 mm$ away from the rod  is shown in 
Fig.~\ref{fig:fig7}(b). It is observed that the dust particles are lifted up a 
height of few mm from the previous equilibrium position (\textit{i.e.}, in 
absence of the rod) adjacent to the floating rod as depicted in 
Fig.~\ref{fig:fig7}(b). It means that the dust particles are now levitated at 
new equilibrium position in presence of the rod and DAWs are originating because 
of collective oscillatory motions of charged dust particles around their 
equilibrium position. In addition, the linear wavefronts of DAW becomes curved 
in nature and found to propagate in Y--Z plane. It is to be noted that the 
average wavelength and phase velocity of DAW are $\sim 2.5~mm$ and $\sim 
3~cm/sec$, respectively. It is worth mentioning that wavelength of DAWs 
increases whereas the velocity decreases when the rod is brought nearer to the 
dust cloud. 
For further characterization of DAWs, the the images are taken in the  horizontal (X--Y) plane. In this plane, the circular wave front of DAWs originates at the outer edge of the perturbed region and propagates in the inward direction (i.e., in the direction of the rod) are observed. A typical video image of DAWs just below the rod in horizontal plane is presented in the Fig.~\ref{fig:fig7}$(c)$. Hence, it can be concluded from (Fig.~\ref{fig:fig7}(b) and Fig.~\ref{fig:fig7}(c)) that the floating rod modifies the propagation characteristics of DAWs which are found to propagates obliquely (along radial ($-\hat{r}$) and gravity ($-\hat{z}$)).
  \par 
It is a fact that the dust grains always follow equipotential contours where the 
electrostatic force acting on them is exactly balanced by the gravitation force. 
In the presence of a floating rod, the cathode sheath electric field gets 
modified and dust grains now follow the modified equipotential contour to get an 
equilibrium position. Hence the dust particles are lifted up a height of few mm 
adjacent to the rod (in perturbed region). In addition, the dynamics of the 
ions gets changed due to the change of sheath electric field. Initially (in 
absence of the rod), the direction of ion streaming is always in $-\hat{z}$ 
direction whereas in presence of the rod the ion flow direction become oblique, 
having velocity components along radial ($-\hat{r}$) and gravity 
($-\hat{z}$)). As it is discussed in earlier section, 
the streaming ions is the main responsible cause to excite the dust acoustic 
waves therefore its propagation direction is changed to oblique. \par
\subsection{Horizontally aligned floating rod}
\begin{figure*}[ht]
 \centering
 \includegraphics[scale=0.850]{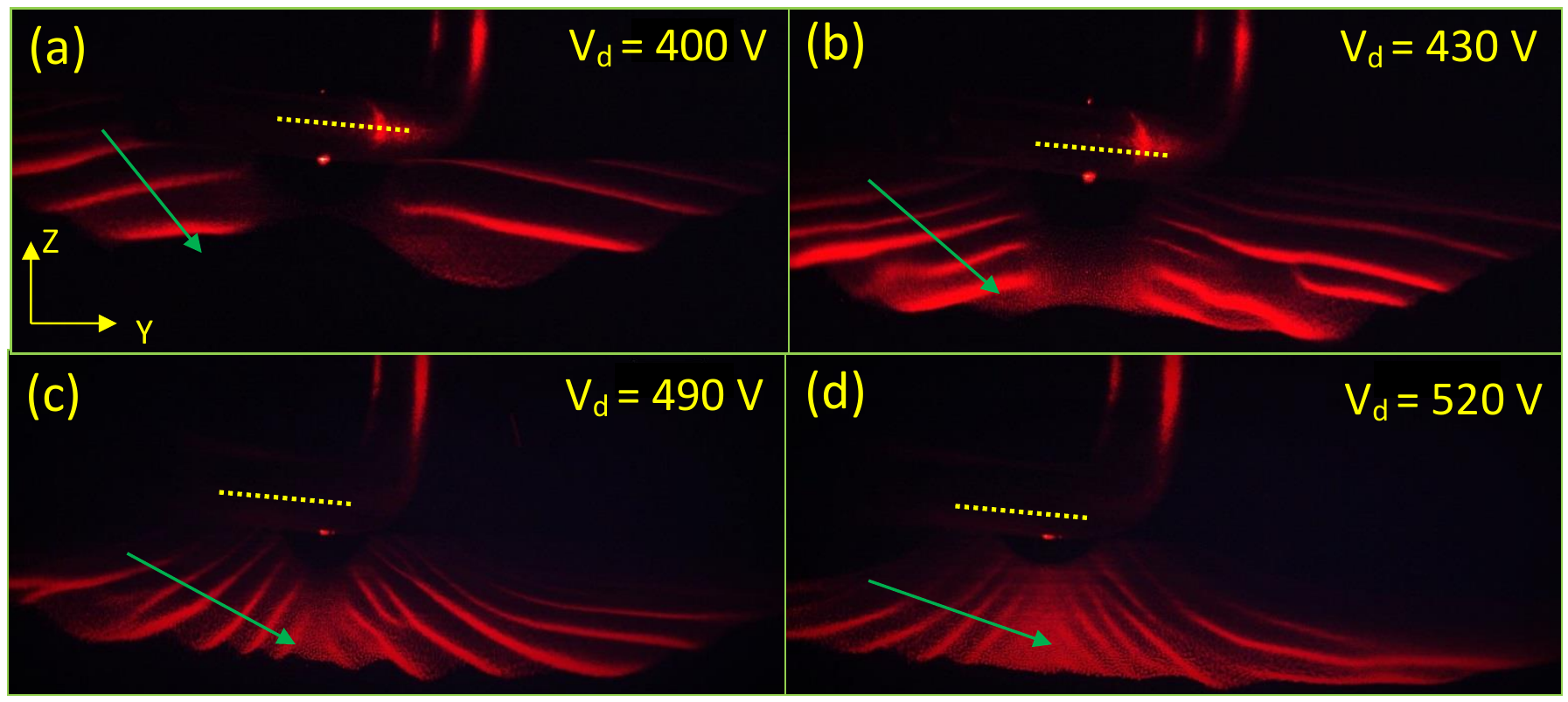} 
  \caption{\label{fig:fig8}Video images of dust $-$ acoustic waves with floating rod in horizontal plane: (a)$-$(d) for discharge voltage $(V_d)$ 400, 430, 490, and 520 V, respectively. The experiments are performed with kaolin particles and gas pressure is set at 0.07 $mbar$. The yellow dotted lines represent the location of floating rod. Green arrow indicates the direction of the velocity vector with respect to direction of gravity.}
 \end{figure*}

To get more insights of the wave$-$rod interaction, a set of experiments is 
performed with a horizontally aligned floating rod which is kept always 
perpendicular to Y--Z plane. To study the propagation characteristics of DAWs 
the images are taken in the Y--Z plane for a particular location of X.  A series 
of experiments have been carried out to study the propagation characteristics of 
DAWs for various discharge voltages $(V_d)$ at a fixed gas pressure (0.07 
$mbar$). It has to be noted that the position of  the floating rod is always 
kept just above the dust cloud (cathode sheath edge) even though the dust cloud 
position changes with the change of discharge voltage. The change of propagation 
characteristics of DAWs are shown in Fig.~\ref{fig:fig8}(a--d) for different 
discharge voltages. In this range of this discharge voltages, the average 
wavelength and phase velocity of DAWs also changes from $\sim 2.6 - 1.7 mm $ and 
$ \sim 3.3 - 4.2 cm/sec$, respectively. These figures clearly indicate the DAWs 
propagates obliquely with two velocity components (along Y and Z) in the 
presence of a rod. The Z$-$component dominates over the Y$-$component at lower 
discharge voltage (400 V) whereas Y$-$component dominates over the Z$-$component 
at higher discharge voltage (520 V). The arrows in Fig.~\ref{fig:fig8}(a) to 
Fig.~\ref{fig:fig8}(d) indicates the propagation dirction of DAWs which is 
obtained by analyzing the consecutive frames for different discharge voltages. 
It is also observed that the dust particles are lifted up a height of few mm 
near the rod and this effect becomes significant for higher discharge voltage 
(Fig.~\ref{fig:fig8}(c) Fig.~\ref{fig:fig8}(d). Similar to earlier case (with 
vertical aligned rod), a dust free region always presents near the rod. The 
volume of the dust free region reduces with the increase of the discharge 
voltage. It is also found that the dust particles just below the rod always 
shows random motion at a discharge voltage of $(380-420 V)$. \par
In the case of horizontally oriented floating rod, sheath around it interacts with cathode sheath. It is expected that modified (resultant) sheath electric field changes with the change of discharge parameters. At lower discharge voltage (400 V), a thick sheath region (strong E--field) around the rod repels the dust grains and dust free region is formed (see Fig.~\ref{fig:fig8}(a)). With the increase of discharge voltage (at 520 V), the sheath thickness around the rod decreases significantly and as a result the void size reduces (see Fig.~\ref{fig:fig8}(d)). The size of dust free region depends on the magnitude of electrostatic and ion--drag force acting on the dust grains.  The size of dust void reduces with increasing the ion--drag force compared to the electrostatic force. A Increase of discharge voltage (from 400 V to 520 V) causes a considerable changes in the sheath electric field which results in a significant change of ion dynamics. It is observed that at lower discharge voltage, the velocity component of ions along Y--direction becomes smaller compare to at higher discharge voltage. As a result the DAWs propagates almost in vertical direction (along the gravity) at lower discharge voltage whereas obliquely at higher discharge voltage. This effect is clearly seen in the Fig.~\ref{fig:fig8}.\par
\begin{figure*}[ht]
 \centering
 \includegraphics[scale=0.85]{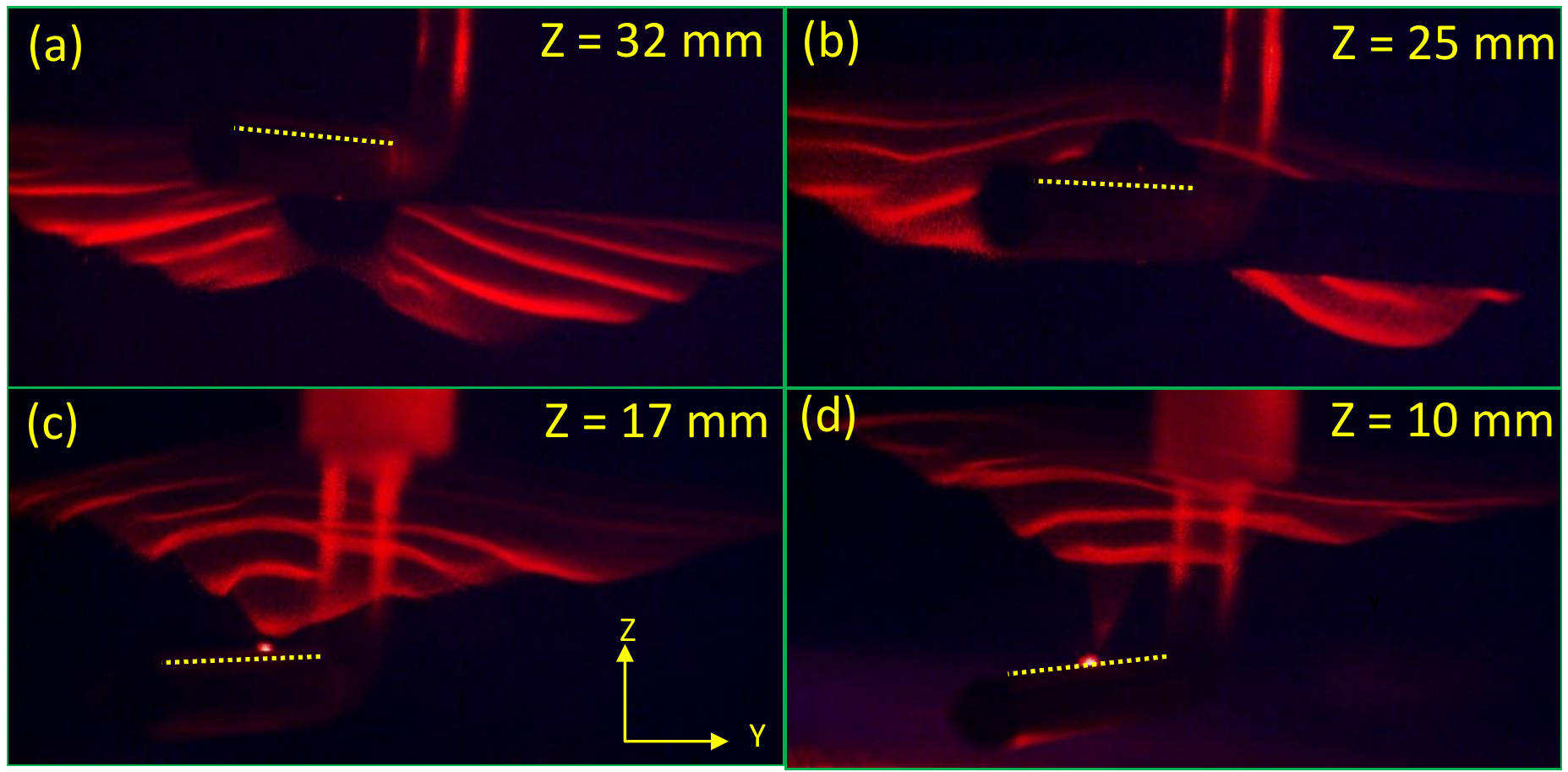} 
 \caption{\label{fig:fig9}Video images of DAWs with floating rod at various locations in the sheath region: (a) rod is about near the edge of presheath or sheath. (b) Rod is inside the sheath region where dust particles are levitated. (c) Rod is just below the lower edge of levitated dust cloud. (d) Rod is near the cathode or inside the sheath. The yellow dotted lines represent the location of rod in this plane. The position of floating rod is measured with respect to the cathode surface. Discharge voltage and pressure are set at 410 V and 0.07 $mbar$.}
 \end{figure*}

 At last, we have further examined the propagation characteristics of DAWs by keeping the floating rod at various Z--locations (measured from cathode surface) for a given discharge condition, $V_d \sim 410$ V and $p \sim 0.07$ mbar. Fig.~\ref{fig:fig9}(a--d) shows the propagation characteristics of DAWs in  presence of this rod for Z $\sim$ 32, 25, 17 and 10 $mm$, respectively. It is to be noted that Z $\sim$ 32$~mm$  corresponds to the location of the topmost layer of the dust cloud. As the floating rod is moved from Z $\sim$ 32 to $25~ mm$, the rod reaches almost to the middle of the dust cloud and modifies the propagation characteristics of DAWs, which is shown in Fig~\ref{fig:fig9}$(b)$. It is seen in Fig.~\ref{fig:fig9}$(b)$, a dust free region of length $\sim$ 2 $mm$ is formed around the cylindrical rod whereas in the outside of the void region, DAW fronts become concave in nature. It is observed that at $Z \sim 17~mm$, the dust cloud takes cone like structure above the floating rod and DAWs propagate obliquely as discussed in earlier section. In the case of Fig~\ref{fig:fig9}$(d)$, the floating rod is moved deeper to the cathode sheath ($Z \sim 10~mm$) where the electric field is most strong compared to three previous locations. In absence of the rod, the dust particle normally could not be found at $\sim 10~mm$ whereas in presence of the rod particles are found to exhibit the oscillatory motion. Interestingly at this location, the dust particles follow the floating rod inside the sheath in the form of a cone like structure which is shown in Fig.~\ref{fig:fig9}$(d)$.  \par
 The observed results with floating rod at different locations provide a strong 
evidence of sheath E$-$field modification around the floating rod while it 
interacts with the dust cloud. { In Fig.~\ref{fig:fig9}$(b)$ (when the rod is 
kept at the edge of cathode sheath), the variation of E--field around the rod is 
similar to that is reported in the ref\citep{Efieldaroundprobe}. The strong E-- 
field near the rod pushes the dust grains against ion drag force and as a result 
dust void is formed. Outside the void region, weak E--field gives rise to the 
curved shape dust cloud (along with the DAWs). In Fig.~\ref{fig:fig9}(c) and 
\ref{fig:fig9}(d), the cone shaped profile of unstable dust cloud is the result 
of modified E$-$field above the rod \citep{Efieldaroundprobe}. The electric 
field gets weaker around the floating rod when it is placed inside the sheath 
region. This lower electric field provides an equllibrium to the dust particles 
above the rod. Dust Acoustic Waves are found to propagate towards the cathode 
inside the confined dust cloud.}
 \section{Summary and conclusions}  \label{sec:conclusion}
 In this paper, we report the experimental observation of self--excited dust 
acoustic waves and its propagation characteristics in the absence and presence of a 
floating cylindrical object. The DC glow discharge dusty plasma is produced in a 
glass tube in the background of argon gas. Plasma and dusty plasma parameters 
are measured/estimated in a wide range of discharge condition. { In our 
experiments, the cathode sheath E--field serves two purposes. Firstly, it 
provides an electrostatic force that balances the dust particles against 
gravity. Secondly, the electric field drives the strong downward ion flow which 
in result excite the DAWs in the dust cloud.}  A thorough investigation on the 
propagation of this dust acoustic wave is carried out with and without a 
floating rod. The main findings of our experimental observations are summarized 
as follows:
 \begin{enumerate}
 \item  Dust acoustic waves are excited spontaneously in an equilibrium dust cloud above a threshold discharge current and below a critical background neutral gas pressure. The low frequency waves are found to propagate along the direction of flow of  ions (towards cathode). A detailed characterisation of DAWs are made by measuring the velocity and the wavelength. 
 \item The propagation characteristics of self--excited DAWs get modified during 
the interaction of a floating cylindrical object either kept vertically or 
horizontally. The modification of DAWs  in presence of the rod is extended up to 
 $> 100\lambda_D$ from the surface of the object. 
 \item In presence of a floating rod (aligned vertically), the DAWs disappear in the perturbed region at lower discharge voltage ($V_d \sim 380-400$ V) whereas at higher discharge voltage (above $V_d \sim 420$ V), the DAW are found to propagate obliquely. The obliqueness of DAWs changes with the change of discharge parameters.
\item In case of horizontally aligned rod, the DAWs gets excited in a cone 
shaped dust cloud in the cathode sheath region. Displacement of dust grains in 
downward (upward) direction confirms that the cathode sheath electric field 
becomes weaker (stronger) in presence of the floating objects. 
\end{enumerate}
The excitation of dust acoustic waves and the modification of its propagation characteristics in presence of a floating object can be explained in terms of streaming ions inside the dust cloud. In absence of the rod, the streaming ions towards the cathode exert a pressure on the levitated dust grains and turns the dust medium into an unstable state, results in the excitation of DAWs which are found to propagate in the direction of gravity. In presence of the rod (kept either vertically or horizontally), the sheath around it alters the potential profile of cathode sheath. It happens due to the coupling between the sheaths formed around the cylindrical rod and the cathode electrode. In this coupling, spatial distribution of electric field gets modified in the perturbed region. In this region, dust grains follow the modified equipotential surfaces to get an equilibrium position therefore they are lifted up near the rod. The direction of flow of ions changes according to the sheath electric field and as a results the DAWs are found to propagate obliquely.  
 
 \section*{Acknowledgement}
 One of the authors M.C. is thankful to Prof. A. Sen, Dr. R. Ganesh and M. 
Sengupta for their valuable suggestion. The authors are greaful to Dr. M. 
Bandyopadhyay for his invaluable inputs to improve the manuscript.
\bibliography{aipsamp}

\begin{thebibliography}{41}%
\makeatletter
\providecommand \@ifxundefined [1]{%
 \@ifx{#1\undefined}
}%
\providecommand \@ifnum [1]{%
 \ifnum #1\expandafter \@firstoftwo
 \else \expandafter \@secondoftwo
 \fi
}%
\providecommand \@ifx [1]{%
 \ifx #1\expandafter \@firstoftwo
 \else \expandafter \@secondoftwo
 \fi
}%
\providecommand \natexlab [1]{#1}%
\providecommand \enquote  [1]{``#1''}%
\providecommand \bibnamefont  [1]{#1}%
\providecommand \bibfnamefont [1]{#1}%
\providecommand \citenamefont [1]{#1}%
\providecommand \href@noop [0]{\@secondoftwo}%
\providecommand \href [0]{\begingroup \@sanitize@url \@href}%
\providecommand \@href[1]{\@@startlink{#1}\@@href}%
\providecommand \@@href[1]{\endgroup#1\@@endlink}%
\providecommand \@sanitize@url [0]{\catcode `\\12\catcode `\$12\catcode
  `\&12\catcode `\#12\catcode `\^12\catcode `\_12\catcode `\%12\relax}%
\providecommand \@@startlink[1]{}%
\providecommand \@@endlink[0]{}%
\providecommand \url  [0]{\begingroup\@sanitize@url \@url }%
\providecommand \@url [1]{\endgroup\@href {#1}{\urlprefix }}%
\providecommand \urlprefix  [0]{URL }%
\providecommand \Eprint [0]{\href }%
\providecommand \doibase [0]{http://dx.doi.org/}%
\providecommand \selectlanguage [0]{\@gobble}%
\providecommand \bibinfo  [0]{\@secondoftwo}%
\providecommand \bibfield  [0]{\@secondoftwo}%
\providecommand \translation [1]{[#1]}%
\providecommand \BibitemOpen [0]{}%
\providecommand \bibitemStop [0]{}%
\providecommand \bibitemNoStop [0]{.\EOS\space}%
\providecommand \EOS [0]{\spacefactor3000\relax}%
\providecommand \BibitemShut  [1]{\csname bibitem#1\endcsname}%
\let\auto@bib@innerbib\@empty
\bibitem [{\citenamefont {Barkan}, \citenamefont {D'Angelo},\ and\
  \citenamefont {Merlino}(1994)}]{Charging}%
  \BibitemOpen
  \bibfield  {author} {\bibinfo {author} {\bibfnamefont {A.}~\bibnamefont
  {Barkan}}, \bibinfo {author} {\bibfnamefont {N.}~\bibnamefont {D'Angelo}}, \
  and\ \bibinfo {author} {\bibfnamefont {R.~L.}\ \bibnamefont {Merlino}},\
  }\bibfield  {title} {\enquote {\bibinfo {title} {Charging of dust grains in a
  plasma},}\ }\href {\doibase 10.1103/PhysRevLett.73.3093} {\bibfield
  {journal} {\bibinfo  {journal} {Phys. Rev. Lett.}\ }\textbf {\bibinfo
  {volume} {73}},\ \bibinfo {pages} {3093--3096} (\bibinfo {year}
  {1994})}\BibitemShut {NoStop}%
\bibitem [{\citenamefont {Rao}, \citenamefont {Shukla},\ and\ \citenamefont
  {Yu}(1990)}]{raodaw1}%
  \BibitemOpen
  \bibfield  {author} {\bibinfo {author} {\bibfnamefont {N.~N.}\ \bibnamefont
  {Rao}}, \bibinfo {author} {\bibfnamefont {P.~K.}\ \bibnamefont {Shukla}}, \
  and\ \bibinfo {author} {\bibfnamefont {M.~Y.}\ \bibnamefont {Yu}},\
  }\bibfield  {title} {\enquote {\bibinfo {title} {Dust-acoustic waves in dusty
  plasmas},}\ }\href@noop {} {\bibfield  {journal} {\bibinfo  {journal}
  {Planet. Space Sci.}\ }\textbf {\bibinfo {volume} {38}},\ \bibinfo {pages}
  {543--546} (\bibinfo {year} {1990})}\BibitemShut {NoStop}%
\bibitem [{\citenamefont {Barkan}, \citenamefont {Merlino},\ and\ \citenamefont
  {D'Angelo}(1995)}]{daw2}%
  \BibitemOpen
  \bibfield  {author} {\bibinfo {author} {\bibfnamefont {A.}~\bibnamefont
  {Barkan}}, \bibinfo {author} {\bibfnamefont {R.~L.}\ \bibnamefont {Merlino}},
  \ and\ \bibinfo {author} {\bibfnamefont {N.}~\bibnamefont {D'Angelo}},\
  }\bibfield  {title} {\enquote {\bibinfo {title} {Laboratory observation of
  the dust-acoustic wave mode},}\ }\href {\doibase
  http://dx.doi.org/10.1063/1.871121} {\bibfield  {journal} {\bibinfo
  {journal} {Phys. Plasmas}\ }\textbf {\bibinfo {volume} {2}},\ \bibinfo
  {pages} {3563--3565} (\bibinfo {year} {1995})}\BibitemShut {NoStop}%
\bibitem [{\citenamefont {Thompson}\ \emph {et~al.}(1997)\citenamefont
  {Thompson}, \citenamefont {Barkan}, \citenamefont {D'Angelo},\ and\
  \citenamefont {Merlino}}]{daw3}%
  \BibitemOpen
  \bibfield  {author} {\bibinfo {author} {\bibfnamefont {C.}~\bibnamefont
  {Thompson}}, \bibinfo {author} {\bibfnamefont {A.}~\bibnamefont {Barkan}},
  \bibinfo {author} {\bibfnamefont {N.}~\bibnamefont {D'Angelo}}, \ and\
  \bibinfo {author} {\bibfnamefont {R.~L.}\ \bibnamefont {Merlino}},\
  }\bibfield  {title} {\enquote {\bibinfo {title} {Dust acoustic waves in a
  direct current glow discharge},}\ }\href {\doibase
  http://dx.doi.org/10.1063/1.872238} {\bibfield  {journal} {\bibinfo
  {journal} {Phys. Plasmas}\ }\textbf {\bibinfo {volume} {4}},\ \bibinfo
  {pages} {2331--2335} (\bibinfo {year} {1997})}\BibitemShut {NoStop}%
\bibitem [{\citenamefont {Schwabe}\ \emph {et~al.}(2007)\citenamefont
  {Schwabe}, \citenamefont {Rubin-Zuzic}, \citenamefont {Zhdanov},
  \citenamefont {Thomas},\ and\ \citenamefont {Morfill}}]{ddw1}%
  \BibitemOpen
  \bibfield  {author} {\bibinfo {author} {\bibfnamefont {M.}~\bibnamefont
  {Schwabe}}, \bibinfo {author} {\bibfnamefont {M.}~\bibnamefont
  {Rubin-Zuzic}}, \bibinfo {author} {\bibfnamefont {S.}~\bibnamefont
  {Zhdanov}}, \bibinfo {author} {\bibfnamefont {H.~M.}\ \bibnamefont {Thomas}},
  \ and\ \bibinfo {author} {\bibfnamefont {G.~E.}\ \bibnamefont {Morfill}},\
  }\bibfield  {title} {\enquote {\bibinfo {title} {Highly resolved self-excited
  density waves in a complex plasma},}\ }\href {\doibase
  10.1103/PhysRevLett.99.095002} {\bibfield  {journal} {\bibinfo  {journal}
  {Phys. Rev. Lett.}\ }\textbf {\bibinfo {volume} {99}},\ \bibinfo {pages}
  {095002} (\bibinfo {year} {2007})}\BibitemShut {NoStop}%
\bibitem [{\citenamefont {Flanagan}\ and\ \citenamefont {Goree}(2010)}]{ddw2}%
  \BibitemOpen
  \bibfield  {author} {\bibinfo {author} {\bibfnamefont {T.~M.}\ \bibnamefont
  {Flanagan}}\ and\ \bibinfo {author} {\bibfnamefont {J.}~\bibnamefont
  {Goree}},\ }\bibfield  {title} {\enquote {\bibinfo {title} {Observation of
  the spatial growth of self-excited dust-density waves},}\ }\href {\doibase
  http://dx.doi.org/10.1063/1.3524691} {\bibfield  {journal} {\bibinfo
  {journal} {Phys. Plasmas}\ }\textbf {\bibinfo {volume} {17}},\ \bibinfo
  {pages} {123702} (\bibinfo {year} {2010})}\BibitemShut {NoStop}%
\bibitem [{\citenamefont {Sarkar}\ \emph {et~al.}(2013)\citenamefont {Sarkar},
  \citenamefont {Bose}, \citenamefont {Mukherjee},\ and\ \citenamefont
  {Pramanik}}]{ddw3}%
  \BibitemOpen
  \bibfield  {author} {\bibinfo {author} {\bibfnamefont {S.}~\bibnamefont
  {Sarkar}}, \bibinfo {author} {\bibfnamefont {M.}~\bibnamefont {Bose}},
  \bibinfo {author} {\bibfnamefont {S.}~\bibnamefont {Mukherjee}}, \ and\
  \bibinfo {author} {\bibfnamefont {J.}~\bibnamefont {Pramanik}},\ }\bibfield
  {title} {\enquote {\bibinfo {title} {Spatiotemporal evolution of dielectric
  driven cogenerated dust density waves},}\ }\href {\doibase
  http://dx.doi.org/10.1063/1.4810804} {\bibfield  {journal} {\bibinfo
  {journal} {Phys. Plasmas}\ }\textbf {\bibinfo {volume} {20}},\ \bibinfo
  {pages} {064502} (\bibinfo {year} {2013})}\BibitemShut {NoStop}%
\bibitem [{\citenamefont {Kaw}\ and\ \citenamefont {Sen}(1998)}]{lmode}%
  \BibitemOpen
  \bibfield  {author} {\bibinfo {author} {\bibfnamefont {P.~K.}\ \bibnamefont
  {Kaw}}\ and\ \bibinfo {author} {\bibfnamefont {A.}~\bibnamefont {Sen}},\
  }\bibfield  {title} {\enquote {\bibinfo {title} {Low frequency modes in
  strongly coupled dusty plasmas},}\ }\href {\doibase
  http://dx.doi.org/10.1063/1.873073} {\bibfield  {journal} {\bibinfo
  {journal} {Phys. Plasmas}\ }\textbf {\bibinfo {volume} {5}},\ \bibinfo
  {pages} {3552--3559} (\bibinfo {year} {1998})}\BibitemShut {NoStop}%
\bibitem [{\citenamefont {Pramanik}\ \emph {et~al.}(2002)\citenamefont
  {Pramanik}, \citenamefont {Prasad}, \citenamefont {Sen},\ and\ \citenamefont
  {Kaw}}]{tsw}%
  \BibitemOpen
  \bibfield  {author} {\bibinfo {author} {\bibfnamefont {J.}~\bibnamefont
  {Pramanik}}, \bibinfo {author} {\bibfnamefont {G.}~\bibnamefont {Prasad}},
  \bibinfo {author} {\bibfnamefont {A.}~\bibnamefont {Sen}}, \ and\ \bibinfo
  {author} {\bibfnamefont {P.~K.}\ \bibnamefont {Kaw}},\ }\bibfield  {title}
  {\enquote {\bibinfo {title} {Experimental observations of transverse shear
  waves in strongly coupled dusty plasmas},}\ }\href {\doibase
  10.1103/PhysRevLett.88.175001} {\bibfield  {journal} {\bibinfo  {journal}
  {Phys. Rev. Lett.}\ }\textbf {\bibinfo {volume} {88}},\ \bibinfo {pages}
  {175001} (\bibinfo {year} {2002})}\BibitemShut {NoStop}%
\bibitem [{\citenamefont {Farokhi}\ \emph {et~al.}(2000)\citenamefont
  {Farokhi}, \citenamefont {Shukla}, \citenamefont {Tsintsadze},\ and\
  \citenamefont {Tskhakaya}}]{dlw1}%
  \BibitemOpen
  \bibfield  {author} {\bibinfo {author} {\bibfnamefont {B.}~\bibnamefont
  {Farokhi}}, \bibinfo {author} {\bibfnamefont {P.~K.}\ \bibnamefont {Shukla}},
  \bibinfo {author} {\bibfnamefont {N.~L.}\ \bibnamefont {Tsintsadze}}, \ and\
  \bibinfo {author} {\bibfnamefont {D.~D.}\ \bibnamefont {Tskhakaya}},\
  }\bibfield  {title} {\enquote {\bibinfo {title} {Dust lattice waves in a
  plasma crystal},}\ }\href {\doibase http://dx.doi.org/10.1063/1.873876}
  {\bibfield  {journal} {\bibinfo  {journal} {Phys. Plasmas}\ }\textbf
  {\bibinfo {volume} {7}},\ \bibinfo {pages} {814--818} (\bibinfo {year}
  {2000})}\BibitemShut {NoStop}%
\bibitem [{\citenamefont {Homann}\ \emph {et~al.}(1998)\citenamefont {Homann},
  \citenamefont {Melzer}, \citenamefont {Peters}, \citenamefont {Madani},\ and\
  \citenamefont {Piel}}]{dlw2}%
  \BibitemOpen
  \bibfield  {author} {\bibinfo {author} {\bibfnamefont {A.}~\bibnamefont
  {Homann}}, \bibinfo {author} {\bibfnamefont {A.}~\bibnamefont {Melzer}},
  \bibinfo {author} {\bibfnamefont {S.}~\bibnamefont {Peters}}, \bibinfo
  {author} {\bibfnamefont {R.}~\bibnamefont {Madani}}, \ and\ \bibinfo {author}
  {\bibfnamefont {A.}~\bibnamefont {Piel}},\ }\bibfield  {title} {\enquote
  {\bibinfo {title} {Laser-excited dust lattice waves in plasma crystals},}\
  }\href {\doibase http://dx.doi.org/10.1016/S0375-9601(98)00141-8} {\bibfield
  {journal} {\bibinfo  {journal} {Physics Letters A}\ }\textbf {\bibinfo
  {volume} {242}},\ \bibinfo {pages} {173 -- 180} (\bibinfo {year}
  {1998})}\BibitemShut {NoStop}%
\bibitem [{\citenamefont {Nunomura}, \citenamefont {Samsonov},\ and\
  \citenamefont {Goree}(2000)}]{tdlw}%
  \BibitemOpen
  \bibfield  {author} {\bibinfo {author} {\bibfnamefont {S.}~\bibnamefont
  {Nunomura}}, \bibinfo {author} {\bibfnamefont {D.}~\bibnamefont {Samsonov}},
  \ and\ \bibinfo {author} {\bibfnamefont {J.}~\bibnamefont {Goree}},\
  }\bibfield  {title} {\enquote {\bibinfo {title} {Transverse waves in a
  two-dimensional screened-coulomb crystal (dusty plasma)},}\ }\href {\doibase
  10.1103/PhysRevLett.84.5141} {\bibfield  {journal} {\bibinfo  {journal}
  {Phys. Rev. Lett.}\ }\textbf {\bibinfo {volume} {84}},\ \bibinfo {pages}
  {5141--5144} (\bibinfo {year} {2000})}\BibitemShut {NoStop}%
\bibitem [{\citenamefont {Rao}(1998)}]{kdv}%
  \BibitemOpen
  \bibfield  {author} {\bibinfo {author} {\bibfnamefont {N.~N.}\ \bibnamefont
  {Rao}},\ }\bibfield  {title} {\enquote {\bibinfo {title} {Dust-acoustic kdv
  solitons in weakly non-ideal dusty plasmas},}\ }\href@noop {} {\bibfield
  {journal} {\bibinfo  {journal} {Phys. Scr.}\ }\textbf {\bibinfo {volume}
  {T75}},\ \bibinfo {pages} {179--181} (\bibinfo {year} {1998})}\BibitemShut
  {NoStop}%
\bibitem [{\citenamefont {Bandyopadhyay}\ \emph {et~al.}(2008)\citenamefont
  {Bandyopadhyay}, \citenamefont {Prasad}, \citenamefont {Sen},\ and\
  \citenamefont {Kaw}}]{pdasw}%
  \BibitemOpen
  \bibfield  {author} {\bibinfo {author} {\bibfnamefont {P.}~\bibnamefont
  {Bandyopadhyay}}, \bibinfo {author} {\bibfnamefont {G.}~\bibnamefont
  {Prasad}}, \bibinfo {author} {\bibfnamefont {A.}~\bibnamefont {Sen}}, \ and\
  \bibinfo {author} {\bibfnamefont {P.~K.}\ \bibnamefont {Kaw}},\ }\bibfield
  {title} {\enquote {\bibinfo {title} {Experimental study of nonlinear dust
  acoustic solitary waves in a dusty plasma},}\ }\href {\doibase
  10.1103/PhysRevLett.101.065006} {\bibfield  {journal} {\bibinfo  {journal}
  {Phys. Rev. Lett.}\ }\textbf {\bibinfo {volume} {101}},\ \bibinfo {pages}
  {065006} (\bibinfo {year} {2008})}\BibitemShut {NoStop}%
\bibitem [{\citenamefont {Nakamura}, \citenamefont {Bailung},\ and\
  \citenamefont {Shukla}(1999)}]{diasw}%
  \BibitemOpen
  \bibfield  {author} {\bibinfo {author} {\bibfnamefont {Y.}~\bibnamefont
  {Nakamura}}, \bibinfo {author} {\bibfnamefont {H.}~\bibnamefont {Bailung}}, \
  and\ \bibinfo {author} {\bibfnamefont {P.~K.}\ \bibnamefont {Shukla}},\
  }\bibfield  {title} {\enquote {\bibinfo {title} {Observation of ion-acoustic
  shocks in a dusty plasma},}\ }\href {\doibase 10.1103/PhysRevLett.83.1602}
  {\bibfield  {journal} {\bibinfo  {journal} {Phys. Rev. Lett.}\ }\textbf
  {\bibinfo {volume} {83}},\ \bibinfo {pages} {1602--1605} (\bibinfo {year}
  {1999})}\BibitemShut {NoStop}%
\bibitem [{\citenamefont {Shukla}\ and\ \citenamefont {Mamun}(2001)}]{dasw}%
  \BibitemOpen
  \bibfield  {author} {\bibinfo {author} {\bibfnamefont {P.~K.}\ \bibnamefont
  {Shukla}}\ and\ \bibinfo {author} {\bibfnamefont {A.~A.}\ \bibnamefont
  {Mamun}},\ }\bibfield  {title} {\enquote {\bibinfo {title} {Dust-acoustic
  shocks in a strongly coupled dusty plasma},}\ }\href {\doibase
  10.1109/27.923698} {\bibfield  {journal} {\bibinfo  {journal} {IEEE Trans.
  Plasma Sci.}\ }\textbf {\bibinfo {volume} {29}},\ \bibinfo {pages} {221--225}
  (\bibinfo {year} {2001})}\BibitemShut {NoStop}%
\bibitem [{\citenamefont {Heinrich}, \citenamefont {Kim},\ and\ \citenamefont
  {Merlino}(2009)}]{expdasw}%
  \BibitemOpen
  \bibfield  {author} {\bibinfo {author} {\bibfnamefont {J.}~\bibnamefont
  {Heinrich}}, \bibinfo {author} {\bibfnamefont {S.-H.}\ \bibnamefont {Kim}}, \
  and\ \bibinfo {author} {\bibfnamefont {R.~L.}\ \bibnamefont {Merlino}},\
  }\bibfield  {title} {\enquote {\bibinfo {title} {Laboratory observations of
  self-excited dust acoustic shocks},}\ }\href {\doibase
  10.1103/PhysRevLett.103.115002} {\bibfield  {journal} {\bibinfo  {journal}
  {Phys. Rev. Lett.}\ }\textbf {\bibinfo {volume} {103}},\ \bibinfo {pages}
  {115002} (\bibinfo {year} {2009})}\BibitemShut {NoStop}%
\bibitem [{\citenamefont {Fortov}\ \emph {et~al.}(2005)\citenamefont {Fortov},
  \citenamefont {Petrov}, \citenamefont {Molotkov}, \citenamefont {Poustylnik},
  \citenamefont {Torchinsky}, \citenamefont {Naumkin},\ and\ \citenamefont
  {Khrapak}}]{exp1dasw}%
  \BibitemOpen
  \bibfield  {author} {\bibinfo {author} {\bibfnamefont {V.~E.}\ \bibnamefont
  {Fortov}}, \bibinfo {author} {\bibfnamefont {O.~F.}\ \bibnamefont {Petrov}},
  \bibinfo {author} {\bibfnamefont {V.~I.}\ \bibnamefont {Molotkov}}, \bibinfo
  {author} {\bibfnamefont {M.~Y.}\ \bibnamefont {Poustylnik}}, \bibinfo
  {author} {\bibfnamefont {V.~M.}\ \bibnamefont {Torchinsky}}, \bibinfo
  {author} {\bibfnamefont {V.~N.}\ \bibnamefont {Naumkin}}, \ and\ \bibinfo
  {author} {\bibfnamefont {A.~G.}\ \bibnamefont {Khrapak}},\ }\bibfield
  {title} {\enquote {\bibinfo {title} {Shock wave formation in a dc glow
  discharge dusty plasma},}\ }\href {\doibase 10.1103/PhysRevE.71.036413}
  {\bibfield  {journal} {\bibinfo  {journal} {Phys. Rev. E}\ }\textbf {\bibinfo
  {volume} {71}},\ \bibinfo {pages} {036413} (\bibinfo {year}
  {2005})}\BibitemShut {NoStop}%
\bibitem [{\citenamefont {Merlino}(2014)}]{dawmerlino}%
  \BibitemOpen
  \bibfield  {author} {\bibinfo {author} {\bibfnamefont {R.~L.}\ \bibnamefont
  {Merlino}},\ }\bibfield  {title} {\enquote {\bibinfo {title} {25 years of
  dust acoustic waves},}\ }\href {\doibase 10.1017/S0022377814000312}
  {\bibfield  {journal} {\bibinfo  {journal} {J. Plasma Physics}\ }\textbf
  {\bibinfo {volume} {80}},\ \bibinfo {pages} {773--786} (\bibinfo {year}
  {2014})}\BibitemShut {NoStop}%
\bibitem [{\citenamefont {Thomas}, \citenamefont {Jr},\ and\ \citenamefont
  {Watson}(1999)}]{dustydevicer}%
  \BibitemOpen
  \bibfield  {author} {\bibinfo {author} {\bibfnamefont {E.}~\bibnamefont
  {Thomas}}, \bibinfo {author} {\bibnamefont {Jr}}, \ and\ \bibinfo {author}
  {\bibfnamefont {M.}~\bibnamefont {Watson}},\ }\bibfield  {title} {\enquote
  {\bibinfo {title} {First experiments in the dusty plasma experiment
  device},}\ }\href {\doibase http://dx.doi.org/10.1063/1.873672} {\bibfield
  {journal} {\bibinfo  {journal} {Phys. Plasmas}\ }\textbf {\bibinfo {volume}
  {6}},\ \bibinfo {pages} {4111--4117} (\bibinfo {year} {1999})}\BibitemShut
  {NoStop}%
\bibitem [{\citenamefont {Trottenberg}, \citenamefont {Block},\ and\
  \citenamefont {Piel}(2006)}]{mddw}%
  \BibitemOpen
  \bibfield  {author} {\bibinfo {author} {\bibfnamefont {T.}~\bibnamefont
  {Trottenberg}}, \bibinfo {author} {\bibfnamefont {D.}~\bibnamefont {Block}},
  \ and\ \bibinfo {author} {\bibfnamefont {A.}~\bibnamefont {Piel}},\
  }\bibfield  {title} {\enquote {\bibinfo {title} {Dust confinement and
  dust-acoustic waves in weakly magnetized anodic plasmas},}\ }\href {\doibase
  http://dx.doi.org/10.1063/1.2196347} {\bibfield  {journal} {\bibinfo
  {journal} {Phys. Plasmas}\ }\textbf {\bibinfo {volume} {13}},\ \bibinfo
  {pages} {042105} (\bibinfo {year} {2006})}\BibitemShut {NoStop}%
\bibitem [{\citenamefont {Molotkov}\ \emph {et~al.}(2004)\citenamefont
  {Molotkov}, \citenamefont {Petrov}, \citenamefont {Pustyl’nik},
  \citenamefont {Torchinskii}, \citenamefont {Fortov},\ and\ \citenamefont
  {Khrapak}}]{ddwfortov}%
  \BibitemOpen
  \bibfield  {author} {\bibinfo {author} {\bibfnamefont {V.~I.}\ \bibnamefont
  {Molotkov}}, \bibinfo {author} {\bibfnamefont {O.~F.}\ \bibnamefont
  {Petrov}}, \bibinfo {author} {\bibfnamefont {M.~Y.}\ \bibnamefont
  {Pustyl’nik}}, \bibinfo {author} {\bibfnamefont {V.~M.}\ \bibnamefont
  {Torchinskii}}, \bibinfo {author} {\bibfnamefont {V.~E.}\ \bibnamefont
  {Fortov}}, \ and\ \bibinfo {author} {\bibfnamefont {A.~G.}\ \bibnamefont
  {Khrapak}},\ }\bibfield  {title} {\enquote {\bibinfo {title} {Dusty plasma of
  a dc glow discharge: methods of investigation and characteristic features of
  behavior},}\ }\href {\doibase http://dx.doi.org/10.1007/s10740-005-0025-4}
  {\bibfield  {journal} {\bibinfo  {journal} {High Temperature}\ }\textbf
  {\bibinfo {volume} {42}},\ \bibinfo {pages} {827--841} (\bibinfo {year}
  {2004})}\BibitemShut {NoStop}%
\bibitem [{\citenamefont {Pramanik}\ \emph {et~al.}(2003)\citenamefont
  {Pramanik}, \citenamefont {Veeresha}, \citenamefont {Prasad},\ and\
  \citenamefont {Sen}}]{pramanikddw}%
  \BibitemOpen
  \bibfield  {author} {\bibinfo {author} {\bibfnamefont {J.}~\bibnamefont
  {Pramanik}}, \bibinfo {author} {\bibfnamefont {B.~M.}\ \bibnamefont
  {Veeresha}}, \bibinfo {author} {\bibfnamefont {G.}~\bibnamefont {Prasad}}, \
  and\ \bibinfo {author} {\bibfnamefont {P.}~\bibnamefont {Sen}, \bibfnamefont
  {A.and~Kaw}},\ }\bibfield  {title} {\enquote {\bibinfo {title} {Experimental
  observation of dust- acoustic wave turbulence},}\ }\href {\doibase
  http://dx.doi.org/10.1016/S0375-9601(03)00614-5} {\bibfield  {journal}
  {\bibinfo  {journal} {Physics Letters A}\ }\textbf {\bibinfo {volume}
  {312}},\ \bibinfo {pages} {84 -- 90} (\bibinfo {year} {2003})}\BibitemShut
  {NoStop}%
\bibitem [{\citenamefont {Chang}\ and\ \citenamefont {Tsai}(2013)}]{defectddw}%
  \BibitemOpen
  \bibfield  {author} {\bibinfo {author} {\bibfnamefont {M.~C.}\ \bibnamefont
  {Chang}}\ and\ \bibinfo {author} {\bibfnamefont {I.}~\bibnamefont {Tsai},
  \bibfnamefont {Y.~Y.and~Lin}},\ }\bibfield  {title} {\enquote {\bibinfo
  {title} {Observation of 3d defect mediated dust acoustic wave turbulence with
  fluctuating defects and amplitude hole filaments},}\ }\href {\doibase
  http://dx.doi.org/10.1063/1.4817802} {\bibfield  {journal} {\bibinfo
  {journal} {Phys. Plasmas}\ }\textbf {\bibinfo {volume} {20}},\ \bibinfo
  {pages} {083703} (\bibinfo {year} {2013})}\BibitemShut {NoStop}%
\bibitem [{\citenamefont {Fortov}\ \emph {et~al.}(2003)\citenamefont {Fortov},
  \citenamefont {Usachev}, \citenamefont {Zobnin}, \citenamefont {Molotkov},\
  and\ \citenamefont {Petrov}}]{icpddw}%
  \BibitemOpen
  \bibfield  {author} {\bibinfo {author} {\bibfnamefont {V.~E.}\ \bibnamefont
  {Fortov}}, \bibinfo {author} {\bibfnamefont {A.~D.}\ \bibnamefont {Usachev}},
  \bibinfo {author} {\bibfnamefont {A.~V.}\ \bibnamefont {Zobnin}}, \bibinfo
  {author} {\bibfnamefont {V.~I.}\ \bibnamefont {Molotkov}}, \ and\ \bibinfo
  {author} {\bibfnamefont {O.~F.}\ \bibnamefont {Petrov}},\ }\bibfield  {title}
  {\enquote {\bibinfo {title} {Dust-acoustic wave instability at the diffuse
  edge of radio frequency inductive low-pressure gas discharge plasma},}\
  }\href {\doibase http://dx.doi.org/10.1063/1.1563667} {\bibfield  {journal}
  {\bibinfo  {journal} {Phys. Plasmas}\ }\textbf {\bibinfo {volume} {10}},\
  \bibinfo {pages} {1199--1208} (\bibinfo {year} {2003})}\BibitemShut {NoStop}%
\bibitem [{\citenamefont {Merlino}(2009{\natexlab{a}})}]{instability1}%
  \BibitemOpen
  \bibfield  {author} {\bibinfo {author} {\bibfnamefont {R.~L.}\ \bibnamefont
  {Merlino}},\ }\bibfield  {title} {\enquote {\bibinfo {title} {Dust-acoustic
  waves driven by an ion-dust streaming instability in laboratory discharge
  dusty plasma experiments},}\ }\href {\doibase
  http://dx.doi.org/10.1063/1.3271155} {\bibfield  {journal} {\bibinfo
  {journal} {Phys. Plasmas}\ }\textbf {\bibinfo {volume} {16}},\ \bibinfo
  {pages} {124501} (\bibinfo {year} {2009}{\natexlab{a}})}\BibitemShut
  {NoStop}%
\bibitem [{\citenamefont {Fortov}\ \emph {et~al.}(2000)\citenamefont {Fortov},
  \citenamefont {Khrapak}, \citenamefont {Khrapak}, \citenamefont {Molotkov},
  \citenamefont {Nefedov}, \citenamefont {Petrov},\ and\ \citenamefont
  {Torchinsky}}]{instability2}%
  \BibitemOpen
  \bibfield  {author} {\bibinfo {author} {\bibfnamefont {V.~E.}\ \bibnamefont
  {Fortov}}, \bibinfo {author} {\bibfnamefont {A.~G.}\ \bibnamefont {Khrapak}},
  \bibinfo {author} {\bibfnamefont {S.~A.}\ \bibnamefont {Khrapak}}, \bibinfo
  {author} {\bibfnamefont {V.~I.}\ \bibnamefont {Molotkov}}, \bibinfo {author}
  {\bibfnamefont {A.~P.}\ \bibnamefont {Nefedov}}, \bibinfo {author}
  {\bibfnamefont {O.~F.}\ \bibnamefont {Petrov}}, \ and\ \bibinfo {author}
  {\bibfnamefont {V.~M.}\ \bibnamefont {Torchinsky}},\ }\bibfield  {title}
  {\enquote {\bibinfo {title} {Mechanism of dust-acoustic instability in a
  direct current glow discharge plasma},}\ }\href@noop {} {\bibfield  {journal}
  {\bibinfo  {journal} {Phys. Plasmas}\ }\textbf {\bibinfo {volume} {7}},\
  \bibinfo {pages} {1374--1380} (\bibinfo {year} {2000})}\BibitemShut {NoStop}%
\bibitem [{\citenamefont {D'Angelo}\ and\ \citenamefont
  {Merlino}(1996)}]{instability3}%
  \BibitemOpen
  \bibfield  {author} {\bibinfo {author} {\bibfnamefont {N.}~\bibnamefont
  {D'Angelo}}\ and\ \bibinfo {author} {\bibfnamefont {R.~L.}\ \bibnamefont
  {Merlino}},\ }\bibfield  {title} {\enquote {\bibinfo {title} {Current driven
  dust acoustic instability in a collisional plasma},}\ }\href {\doibase
  http://dx.doi.org/10.1016/S0032-0633(96)00069-4} {\bibfield  {journal}
  {\bibinfo  {journal} {Planet. Space Sci.}\ }\textbf {\bibinfo {volume}
  {44}},\ \bibinfo {pages} {1593 --1598} (\bibinfo {year} {1996})}\BibitemShut
  {NoStop}%
\bibitem [{\citenamefont {Mamun}\ and\ \citenamefont
  {Shukla}(2000)}]{instability4}%
  \BibitemOpen
  \bibfield  {author} {\bibinfo {author} {\bibfnamefont {A.~A.}\ \bibnamefont
  {Mamun}}\ and\ \bibinfo {author} {\bibfnamefont {P.~K.}\ \bibnamefont
  {Shukla}},\ }\bibfield  {title} {\enquote {\bibinfo {title} {Streaming
  instabilities in a collisional dusty plasma},}\ }\href {\doibase
  http://dx.doi.org/10.1063/1.1315305} {\bibfield  {journal} {\bibinfo
  {journal} {Phys. Plasmas}\ }\textbf {\bibinfo {volume} {7}},\ \bibinfo
  {pages} {4412--4417} (\bibinfo {year} {2000})}\BibitemShut {NoStop}%
\bibitem [{\citenamefont {Rosenberg}(2002)}]{instability5}%
  \BibitemOpen
  \bibfield  {author} {\bibinfo {author} {\bibfnamefont {M.}~\bibnamefont
  {Rosenberg}},\ }\bibfield  {title} {\enquote {\bibinfo {title} {A note on
  ion–dust streaming instability in a collisional dusty plasma},}\ }\href
  {\doibase 10.1017/S0022377802001678} {\bibfield  {journal} {\bibinfo
  {journal} {J. Plasma Physics}\ }\textbf {\bibinfo {volume} {67}},\ \bibinfo
  {pages} {235--242} (\bibinfo {year} {2002})}\BibitemShut {NoStop}%
\bibitem [{\citenamefont {Thompson}, \citenamefont {D’Angelo},\ and\
  \citenamefont {Merlino}(1999)}]{moveprobe}%
  \BibitemOpen
  \bibfield  {author} {\bibinfo {author} {\bibfnamefont {C.~O.}\ \bibnamefont
  {Thompson}}, \bibinfo {author} {\bibfnamefont {N.}~\bibnamefont
  {D’Angelo}}, \ and\ \bibinfo {author} {\bibfnamefont {R.~L.}\ \bibnamefont
  {Merlino}},\ }\bibfield  {title} {\enquote {\bibinfo {title} {The interaction
  of stationary and moving objects with dusty plasmas},}\ }\href {\doibase
  http://dx.doi.org/10.1063/1.873392} {\bibfield  {journal} {\bibinfo
  {journal} {Phys. Plasmas}\ }\textbf {\bibinfo {volume} {6}},\ \bibinfo
  {pages} {1421--1426} (\bibinfo {year} {1999})}\BibitemShut {NoStop}%
\bibitem [{\citenamefont {Thomas}\ \emph {et~al.}(2004)\citenamefont {Thomas},
  \citenamefont {Jr.}, \citenamefont {Avinash},\ and\ \citenamefont
  {Merlino}}]{void}%
  \BibitemOpen
  \bibfield  {author} {\bibinfo {author} {\bibfnamefont {E.}~\bibnamefont
  {Thomas}}, \bibinfo {author} {\bibnamefont {Jr.}}, \bibinfo {author}
  {\bibfnamefont {K.}~\bibnamefont {Avinash}}, \ and\ \bibinfo {author}
  {\bibfnamefont {R.~L.}\ \bibnamefont {Merlino}},\ }\bibfield  {title}
  {\enquote {\bibinfo {title} {Probe induced voids in a dusty plasma},}\ }\href
  {\doibase http://dx.doi.org/10.1063/1.1688333} {\bibfield  {journal}
  {\bibinfo  {journal} {Phys. Plasmas}\ }\textbf {\bibinfo {volume} {11}},\
  \bibinfo {pages} {1770--1774} (\bibinfo {year} {2004})}\BibitemShut {NoStop}%
\bibitem [{\citenamefont {Sarkar}\ \emph {et~al.}(2015)\citenamefont {Sarkar},
  \citenamefont {Mondal}, \citenamefont {Bose},\ and\ \citenamefont
  {Mukherjee}}]{ringvoid}%
  \BibitemOpen
  \bibfield  {author} {\bibinfo {author} {\bibfnamefont {S.}~\bibnamefont
  {Sarkar}}, \bibinfo {author} {\bibfnamefont {M.}~\bibnamefont {Mondal}},
  \bibinfo {author} {\bibfnamefont {M.}~\bibnamefont {Bose}}, \ and\ \bibinfo
  {author} {\bibfnamefont {S.}~\bibnamefont {Mukherjee}},\ }\bibfield  {title}
  {\enquote {\bibinfo {title} {Observation of external control and formation of
  a void in cogenerated dusty plasma},}\ }\href {\doibase
  http://dx.doi.org/10.1088/0963-0252/24/3/035007} {\bibfield  {journal}
  {\bibinfo  {journal} {Plasma Sources Sci. Technol.}\ }\textbf {\bibinfo
  {volume} {24}},\ \bibinfo {pages} {035007} (\bibinfo {year}
  {2015})}\BibitemShut {NoStop}%
\bibitem [{\citenamefont {Klindworth}\ \emph {et~al.}(2004)\citenamefont
  {Klindworth}, \citenamefont {Piel}, \citenamefont {Melzer}, \citenamefont
  {Konopka}, \citenamefont {Rothermel}, \citenamefont {Tarantik},\ and\
  \citenamefont {Morfill}}]{microgravityvoid}%
  \BibitemOpen
  \bibfield  {author} {\bibinfo {author} {\bibfnamefont {M.}~\bibnamefont
  {Klindworth}}, \bibinfo {author} {\bibfnamefont {A.}~\bibnamefont {Piel}},
  \bibinfo {author} {\bibfnamefont {A.}~\bibnamefont {Melzer}}, \bibinfo
  {author} {\bibfnamefont {U.}~\bibnamefont {Konopka}}, \bibinfo {author}
  {\bibfnamefont {H.}~\bibnamefont {Rothermel}}, \bibinfo {author}
  {\bibfnamefont {K.}~\bibnamefont {Tarantik}}, \ and\ \bibinfo {author}
  {\bibfnamefont {G.~E.}\ \bibnamefont {Morfill}},\ }\bibfield  {title}
  {\enquote {\bibinfo {title} {Dust-free regions around langmuir probes in
  complex plasmas under microgravity},}\ }\href {\doibase
  10.1103/PhysRevLett.93.195002} {\bibfield  {journal} {\bibinfo  {journal}
  {Phys. Rev. Lett.}\ }\textbf {\bibinfo {volume} {93}},\ \bibinfo {pages}
  {195002} (\bibinfo {year} {2004})}\BibitemShut {NoStop}%
\bibitem [{\citenamefont {Kim}, \citenamefont {Heinrich},\ and\ \citenamefont
  {Merlino}(2008)}]{diffraction}%
  \BibitemOpen
  \bibfield  {author} {\bibinfo {author} {\bibfnamefont {S.-H.}\ \bibnamefont
  {Kim}}, \bibinfo {author} {\bibfnamefont {J.~R.}\ \bibnamefont {Heinrich}}, \
  and\ \bibinfo {author} {\bibfnamefont {R.~L.}\ \bibnamefont {Merlino}},\
  }\bibfield  {title} {\enquote {\bibinfo {title} {Diffraction of dust acoustic
  waves by a circular cylinder},}\ }\href {\doibase
  http://dx.doi.org/10.1063/1.2977986} {\bibfield  {journal} {\bibinfo
  {journal} {Phys. Plasmas}\ }\textbf {\bibinfo {volume} {15}},\ \bibinfo
  {pages} {090701} (\bibinfo {year} {2008})}\BibitemShut {NoStop}%
\bibitem [{\citenamefont {Meyer}\ \emph {et~al.}(2013)\citenamefont {Meyer},
  \citenamefont {Heinrich}, \citenamefont {Kim},\ and\ \citenamefont
  {Merlino}}]{flowing}%
  \BibitemOpen
  \bibfield  {author} {\bibinfo {author} {\bibfnamefont {J.~K.}\ \bibnamefont
  {Meyer}}, \bibinfo {author} {\bibfnamefont {J.~R.}\ \bibnamefont {Heinrich}},
  \bibinfo {author} {\bibfnamefont {S.-H.}\ \bibnamefont {Kim}}, \ and\
  \bibinfo {author} {\bibfnamefont {R.~L.}\ \bibnamefont {Merlino}},\
  }\bibfield  {title} {\enquote {\bibinfo {title} {Interaction of a biased
  cylinder with a flowing dusty plasma},}\ }\href {\doibase
  10.1017/S0022377813000299} {\bibfield  {journal} {\bibinfo  {journal} {J.
  Plasma Physics}\ }\textbf {\bibinfo {volume} {79}},\ \bibinfo {pages}
  {677--682} (\bibinfo {year} {2013})}\BibitemShut {NoStop}%
\bibitem [{\citenamefont {Law}\ \emph {et~al.}(1998)\citenamefont {Law},
  \citenamefont {Steel}, \citenamefont {Annaratone},\ and\ \citenamefont
  {Allen}}]{circulation}%
  \BibitemOpen
  \bibfield  {author} {\bibinfo {author} {\bibfnamefont {D.~A.}\ \bibnamefont
  {Law}}, \bibinfo {author} {\bibfnamefont {W.~H.}\ \bibnamefont {Steel}},
  \bibinfo {author} {\bibfnamefont {B.~M.}\ \bibnamefont {Annaratone}}, \ and\
  \bibinfo {author} {\bibfnamefont {J.~E.}\ \bibnamefont {Allen}},\ }\bibfield
  {title} {\enquote {\bibinfo {title} {Probe-induced particle circulation in a
  plasma crystal},}\ }\href {\doibase 10.1103/PhysRevLett.80.4189} {\bibfield
  {journal} {\bibinfo  {journal} {Phys. Rev. Lett.}\ }\textbf {\bibinfo
  {volume} {80}},\ \bibinfo {pages} {4189--4192} (\bibinfo {year}
  {1998})}\BibitemShut {NoStop}%
\bibitem [{\citenamefont {Choudhary}, \citenamefont {Mukherjee},\ and\
  \citenamefont {Bandyopadhyay}()}]{expsystem}%
  \BibitemOpen
  \bibfield  {author} {\bibinfo {author} {\bibfnamefont {M.}~\bibnamefont
  {Choudhary}}, \bibinfo {author} {\bibfnamefont {S.}~\bibnamefont
  {Mukherjee}}, \ and\ \bibinfo {author} {\bibfnamefont {P.}~\bibnamefont
  {Bandyopadhyay}},\ }\bibfield  {title} {\enquote {\bibinfo {title} {Transport
  and trapping of dust particles in a potential well created by inductively
  coupled diffused plasma.}}\ }\href@noop {} {\bibinfo  {journal}
  {arXiv:1604.03368}\ }\BibitemShut {NoStop}%
\bibitem [{\citenamefont {Sheridan}\ and\ \citenamefont
  {Goree}(1991)}]{sheaththickness}%
  \BibitemOpen
\bibfield  {journal} {  }\bibfield  {author} {\bibinfo {author} {\bibfnamefont
  {T.~E.}\ \bibnamefont {Sheridan}}\ and\ \bibinfo {author} {\bibfnamefont
  {J.}~\bibnamefont {Goree}},\ }\bibfield  {title} {\enquote {\bibinfo {title}
  {Collisional plasma sheath model},}\ }\href {\doibase
  http://dx.doi.org/10.1063/1.859987} {\bibfield  {journal} {\bibinfo
  {journal} {Phys.Fluids B}\ }\textbf {\bibinfo {volume} {3}},\ \bibinfo
  {pages} {2796--2804} (\bibinfo {year} {1991})}\BibitemShut {NoStop}%
\bibitem [{\citenamefont {Merlino}(2009{\natexlab{b}})}]{sounddaw}%
  \BibitemOpen
  \bibfield  {author} {\bibinfo {author} {\bibfnamefont {R.~L.}\ \bibnamefont
  {Merlino}},\ }\bibfield  {title} {\enquote {\bibinfo {title} {Dust-acoustic
  waves: Visible sound waves},}\ }\href@noop {} {\bibfield  {journal} {\bibinfo
   {journal} {AIP Conf. Proc.}\ }\textbf {\bibinfo {volume} {1188}},\ \bibinfo
  {pages} {141--152} (\bibinfo {year} {2009}{\natexlab{b}})}\BibitemShut
  {NoStop}%
\bibitem [{\citenamefont {Barnat}\ and\ \citenamefont
  {Hebner}(2007)}]{Efieldaroundprobe}%
  \BibitemOpen
  \bibfield  {author} {\bibinfo {author} {\bibfnamefont {E.~V.}\ \bibnamefont
  {Barnat}}\ and\ \bibinfo {author} {\bibfnamefont {G.~A.}\ \bibnamefont
  {Hebner}},\ }\bibfield  {title} {\enquote {\bibinfo {title} {Electric field
  profiles around an electrical probe immersed in a plasma},}\ }\href {\doibase
  http://dx.doi.org/10.1063/1.2404471} {\bibfield  {journal} {\bibinfo
  {journal} {J. Appl. Phys.}\ }\textbf {\bibinfo {volume} {101}},\ \bibinfo
  {pages} {013306} (\bibinfo {year} {2007})}\BibitemShut {NoStop}%
\end{thebibliography}%
\end{document}